\documentclass[11pt,a4paper]{article}
\pdfoutput=1
\usepackage{ifpdf}
\usepackage{jheppub}
\usepackage{rotating}
\usepackage[bb=boondox]{mathalfa}
\usepackage{comment}
\usepackage{tikz-cd}

\graphicspath{ {./figures/} }
\usepackage{mathrsfs}
\usepackage{amsmath,amsfonts,amssymb}
\usepackage{mathtools}
\usepackage{dcolumn}
\usepackage{bm}
\usepackage[mathlines]{lineno}
\usepackage{float}
\usepackage{blindtext}
\usepackage{titlesec}
\title{Sections and Chapters}
\usepackage[toc,page]{appendix}
\usepackage{cancel}
\usepackage[thinc]{esdiff}
\usepackage[euler]{textgreek}
\usepackage{braket}
\usepackage{physics}
\usepackage{xfrac}
\usepackage{soul}
\usepackage{stmaryrd}
\usepackage{pifont}
\usepackage{subfigure}
\usepackage{gensymb}
\usepackage{makecell}
\usepackage{orcidlink}
\usepackage{slashed}
\usepackage{multirow}
\usepackage[normalem]{ulem}
\usepackage{booktabs}

\usepackage{tikz}
\usepackage[compat=1.1.0]{tikz-feynman}

\usepackage[compat=1.1.0]{tikz-feynman}
\usetikzlibrary{positioning,arrows}
\usetikzlibrary{decorations.pathmorphing}
\usetikzlibrary{decorations.markings}

\usepackage{cancel}
\usepackage[normalem]{ulem}

\definecolor{darkorange}{cmyk}{0,0.,1,0.4}

\definecolor{darkgreen}{cmyk}{1,0,1,0.4}

\def\beq{\begin{equation}}
\def\eeq{\end{equation}}
\def\barr{\begin{array}}
\def\earr{\end{array}}
\def\dis{\displaystyle}
\def\lb{\mathfrak{L}_\beta}

\newcommand{\lsim}{
\mathrel{\hbox{\rlap{\hbox{\lower4pt\hbox{$\sim$}}}\hbox{$<$}}}}

\newcommand{\gsim}{
\mathrel{\hbox{\rlap{\hbox{\lower4pt\hbox{$\sim$}}}\hbox{$>$}}}}

\renewcommand{\arraystretch}{2}

\allowdisplaybreaks[2]

\newcommand{\nn}{\nonumber}

\def\re{{\rm Re}}

\definecolor{schrift}{RGB}{120,0,0}

\title{\boldmath{Soft photon corrections in  $B \to K^{(\ast)} \ell^+ \ell^-$ and $\Lambda_b \to \Lambda^{(\ast)} \ell^+ \ell^-$ decays}}

\author[a]{Debajyoti Choudhury,\,\orcidlink{0000-0002-8124-0043}\,}
\author[a,b]{Diganta Das,\,\orcidlink{0000-0003-4570-2738}\,}
\author[a]{Jaydeb Das,\,\orcidlink{0000-0001-6335-9377}\,}

\emailAdd{debchou.physics@gmail.com}
\emailAdd{diganta.das@iiit.ac.in}
\emailAdd{jaydebphysics@gmail.com}

\affiliation[a]{Department of Physics and Astrophysics, University of
  Delhi, Delhi 110007, India} \affiliation[b]{Center for Computational
  Natural Sciences and Bioinformatics, International Institute of
  Information Technology, Hyderabad 500 032, India}

\abstract{We calculate QED corrections to the semileptonic decays $H_1
  \to H_2\ell^+\ell^-$ where $\ell=e, \mu$ and $H_{1,2}$ are
  hadrons. The soft and/or collinear divergences are regulated in a
  gauge-invariant manner and demonstrably cancel, leaving behind a
  finite residue that depends on the (infrared) momentum cutoff below
  which a photon is considered to be indistinguishable. On resumming,
  the said sensitivity reduces drastically, {\em i.e.}, for the NLL
  result as compared to the NLO one. The overall correction is
  negative and its magnitude is larger for a lighter
  lepton. The corrections are presented for $B\to K^{(*)}$ and, for the first time, for $\Lambda_b\to\Lambda^{(*)}$ decays. Given that QED final-state radiation is routinely incorporated in experimental Monte Carlo simulations, for example through PHOTOS, our results provide an important theoretical framework for assessing the accuracy and limitations of these simulations and their impact on LFU observables such as $R_{K^{(*)}}$. }


\begin{document}

\tikzset{
fermion/.style={solid,draw=black, postaction={decorate},decoration={markings,mark=at position 0.8 with {\arrow[draw=black,thick]{>}}}},
antifermion/.style={solid,draw=black, postaction={decorate},decoration={markings,mark=at position 0.8 with {\arrow[draw=black,thick]{<}}}},
majoranafermion/.style={solid,draw=black, postaction={decorate},decoration={markings,mark=at position 0.5 with {\arrow[draw=black,thick]{><}}}},
vector/.style={decorate, decoration={snake, amplitude=0.8mm, segment length=2mm, post length=0.1mm}, draw},
vectorbend/.style={decorate, decoration={snake, amplitude=0.8mm, segment length=2mm, post length=0.1mm,bend left}, draw}
}

\maketitle

\renewcommand{\arraystretch}{1.6}

\section{Introduction}
The flavor changing neutral current
transition $b\to s\ell^+\ell^-$ is, within the
Standard Model (SM), not only loop suppressed, but also
suppressed by factors of quark mixings (CKM suppression). It is, thus,
possible that small contributions from putative New Physics (NP)
sources may no longer be overwhelmed by their SM counterparts,
rendering this sector a promising theatre for searching for such NP
effects. Of late, a particular aspect, namely lepton flavor
universality (LFU) has attracted much attention. Within the SM, a
violation of LFU can be generated only through Higgs-mediated
amplitudes, which, in any case, are enormously suppressed on account
of the smallness of the Yukawa couplings. Such a universality, of
course, need not be respected in NP-mediated
contributions. Observables probing LFU are aplenty, for example the
$B\to K^{(\ast)}\ell^+\ell^-$ transitions.  In particular, the ratios 
\begin{equation}
\label{eq:RK}
R_{K^{(*)}} \equiv \frac{ {\rm BR}(B\to K^{(*)} \mu^+\mu^-)}{ {\rm BR}(B\to K^{(*)}e^+ e^-)}\,
\end{equation}
are expected to be largely free from uncertainties in the hadronic
matrix element and to differ from unity only an account of the
relatively small effect of the lepton masses in the kinematic
factors.  A series of experiments (at BaBar, Belle and LHCb
{\color{blue}\cite{LHCb:2017avl,Belle:2009zue,LHCb:2013ghj,LHCb:2014vgu}})
had indicated a significant discrepancy in $R_{K^{\ast}}$.  However, the most recent
results on $R_{K^{(\ast)}}$~\cite{LHCb:2022zom} namely, 
\begin{eqnarray} \label{data:lowqsq}
 \text{Low-}q^2 &:&
\left\{
	\begin{array}{ll}
		R_{K^\ast}  = 0.994^{+0.090}_{-0.082} \text{ (stat)} {}^{+0.029}_{-0.027} \text{ (syst)} \\
		R_{K}  = 0.927^{+0.093}_{-0.087} \text{ (stat)} {}^{+0.036}_{-0.035} \text{ (syst)}
	\end{array}
        \right.\\
        \label{data:centqsq}
\text{Central-}q^2 &:&
\left\{
	\begin{array}{ll}
		R_{K^\ast}  = 0.949^{+0.042}_{-0.041} \text{ (stat)} {}^{+0.022}_{-0.022} \text{ (syst)} \\
		R_{K}  = 1.027^{+0.072}_{-0.068} \text{ (stat)} {}^{+0.027}_{-0.026} \text{ (syst)}
	\end{array}
\right. \ ,
\end{eqnarray}
indicate a consistency with the SM at $1.2\sigma$. In spite of this,
there is still room for significant LFU violation , should the
NP be CP
violating ~\cite{Fleischer:2023zeo}.

The question of LFU violation aside, in the neutral current sector,
there exist many longstanding hints of NP in the $b\to s\ell^+\ell^-$
transitions angular observables. For example, there is a long standing
discrepancy in the $P_5^\prime$ observable in the $B\to
K^\ast\mu^+\mu^-$ transition
\cite{LHCb:2013ghj,LHCb:2015svh,LHCb:2020lmf}, and systematic
deficits in\footnote{Note that this deficit,
  alongwith the latest result on $R_{K^{(\ast)}}$ would indicate a
  deficit in the differential rates of electron modes as well. In
  other words, there is a case for at least a flavour-universal
  NP.}  both $B\to K^{(\ast)}\mu^+\mu^-$, $B_s\to
\phi\mu^+\mu^-$ \cite{LHCb:2015wdu,LHCb:2021xxq}.  Given the plethora
of discrepancies, further investigation of the $b\to s\ell^+\ell^-$
seems to be warranted. The $b$-flavored baryon $\Lambda_b$ provides
one such avenue. Recently, the LHCb presented data on the differential
decay widths including angular observables for both
$\Lambda_b\to \Lambda\mu^+\mu^-$ \cite{LHCb:2015tgy,LHCb:2023ptw} and
$\Lambda_b\to \Lambda^\ast \mu^+\mu^-$ \cite{LHCb:2023ptw} decays.  It
is observed that, in the $\Lambda_b\to \Lambda\mu^+\mu^-$ differential
decay rates \cite{LHCb:2015tgy}, there is a deficit at low dilepton
invariant masses, but an excess at high dilepton masses.  On the other
hand, the LHCb also performed a test of LFU in $\Lambda_b\to
pK^-\ell^+\ell^-$ \cite{LHCb:2019efc} and found it to be consistent
with the SM expectations.

Mention must also be made of analogous
deviations~\cite{Belle:2015qfa,BaBar:2012obs,BaBar:2013mob,Belle:2016kgw,LHCb:2015gmp,Abdesselam:2016xqt,HFLAV:2014fzu} from
theoretical expectations in corresponding charged current
interactions, such as
\[
R(D^{(*)}) \equiv \frac{ {\rm BR}(B\to D^{(*)}\tau\nu)}{ {\rm BR}(B\to D^{(*)}\ell \nu)}\, , \qquad \quad
R_{J/\psi} \equiv \frac{ {\rm BR}(B_c\to J/\psi\, \tau\nu)}{ {\rm BR}(B_c\to J/\psi\, \mu \nu)}\,
\]
(with $\ell = e, \mu$). With the experimental difficulty in measuring
modes with neutrinos in the final state being more than compensated by
the much higher branching fractions, the initially observed large
deviations had led to sustained exploration of possible scenarios that
could address these anomalies either singly or together.  Recently
measured values, namely $R(D^\ast) = 0.281 \pm 0.018 \pm 0.024$
\cite{LHCb:2023zxo} and $R(D) = 0.441 \pm 0.060 \pm 0.066 $
\cite{LHCb:2023zxo} (where the first uncertainty is statistical and
the second is systematic), though, indicate only $1.9\sigma$
deviations from the SM predictions. Similarly, the
  result for $R_{J/\Psi} (= 0.71 \pm 0.17 \pm 0.18)$
  \cite{LHCb:2017vlu} lies nearly $2\sigma$ above the SM
  expectations. In other words, while the individual observable in the
  charged current sector does not call for new physics, together they
  represent a tension with the SM.

If the aforementioned discrepancies in decay rates are confirmed,
these could constitute signals for NP. Unfortunately, differential
decay rates are not as clean as the LFU ratios. It is therefore
imperative to have control over the theoretical uncertainties. The
form factors, parametrizing the hadronic matrix elements, constitute
the dominant sources of theoretical uncertainties. Such uncertainties
are expected to reduce drastically in future lattice QCD
calculations. Far less attention has been paid to the QED corrections
that too need to be systematically included. However, such QED corrections to $B\to K\ell^+\ell^-$ have been studied in Refs.~\cite{Isidori:2020acz,Mishra:2020orb}, while their implications for the LFU ratios $R_K$ and $R_{K^\ast}$ were investigated in Ref.~\cite{Bordone:2016gaq}. Furthermore, these corrections have been compared with the PHOTOS~\cite{Barberio:1993qi,Barberio:1990ms} Monte Carlo used in experimental analyses, with good agreement found for the neutral $B\to K\ell^+\ell^-$ channel~\cite{Isidori:2022bzw}. In the present work, we extend these studies by providing a systematic analysis of QED corrections to the differential decay distributions of $B\to K\ell^+\ell^-$ and $B\to K^\ast\ell^+\ell^-$, as well as the baryonic decays $\Lambda_b\to\Lambda^{(*)}\ell^+\ell^-$. We also examine how next-to-leading logarithmic (NLL) resummation reduces the dependence of the corrections on the infrared momentum cutoff. In addition, we also consider, for
the first time, $\Lambda_b \to \Lambda^{(\ast)}\ell^+\ell^-$ baryonic
modes. We include both real soft photon emissions as well as virtual
(self-energy and vertex) corrections. Since the initial and the final
hadrons are neutral, to the lowest non-trivial order, photons may be
considered to be emitted from the charged leptons alone.
Contributions emanating from photon emissions from the hadrons (owing
to their inherent structure) are free of infrared
  divergences. Moreover, for the non-helicity-suppressed transitions
  considered here, they do not generate additional hard-collinear
  logarithms~\cite{Isidori:2020acz,
    Nabeebaccus:2022pje,Zwicky:2021olr} (see ref.\cite{Rowe:2024jml}
  for helicity-suppressed modes). We therefore neglect them, and work
  in the point-like approximation.

To $\mathcal{O}(\alpha_{\rm em})$, {\em i.e.}, the lowest order in the
QED corrections, the four-body $H_1(p_0)\to
H_2(p_1)\ell^-(q_1)\ell^+(q_2)\gamma(k)$ decay amplitude splits into
two individually gauge-invariant subamplitudes: a term analogous to
Low's soft photon amplitude that consists of an amplitude for the
non-radiative process and a universal function that is IR divergent,
and a finite part. We use a nonzero photon mass as a regulator for the
infrared divergences.  It is explicitly checked that all dependences
on this fictitious mass cancels once all the contributions, virtual as
well as real emissions, are added. Additionally, a minimal photon
momentum is introduced as a cut-off so as to account for the fact
that, in actual experiments, photons below a certain momentum are not
detectable. In other words, this ``IR cut-off'' defines the boundary
for what is to be considered as a correction and what constitutes an
observable state with an extra photon. Understandably, at the
differential level, the results are sensitive to this cut-off scale,
with the dependence reducing as even higher order terms are
included. The overall correction is found to be negative, and the
numerical value is larger for smaller lepton mass. Beyond
$\mathcal{O}(\alpha_{\rm em})$, the effect of infinite number of
photon emission is included by exponentiating the lowest order
correction, following ref.\cite{Yennie:1961ad} where it is shown that
the cancellation of infrared divergence is similar to that in the
lowest order. The overall corrections being negative, we find that in
the differential branching ratio for the electrons mode, the
corrections are in the 8--18\% range at low-$q^2$ and between 18--30\%
for the high-$q^2$ region\footnote{The exact sizes of the corrections
depend, of course, on the order of the calculation ({\em e.g.}, NLO
versus NLL) and the value of the aforementioned IR cut-off.}. On the
other hand, for the muonic mode, the corrections are 1--4\% in
low-$q^2$ region and 5--10\% in the high-$q^2$
region. Although, relative to a non-radiative
  computation, these corrections enhance the LFU ratios $R_{K^{(*)}}$,
  this should not be interpreted as reviving the tension in these
  observables, since the experimental analyses have already accounted
  for such radiation using PHOTOS.

The paper is organized as follows: in section \ref{sec:nonrad}, we briefly review the formalism of non-radiative decay. In section \ref{sec:qed}, we discuss the structure of the QED corrections ({\em i.e.}, both soft photon emissions and one-loop virtual corrections). Following that, we provide a numerical analysis in section \ref{sec:obsv}. A summary of our work is provided in section \ref{sec:summary}.

\section{Non-radiative decay}\label{sec:nonrad}
Within the SM, the $b\to s \ell^+\ell^-$ transitions proceed, to the
leading order, through electroweak penguin and box diagrams. The
corresponding Hamiltonian is dominated by three operators, {\em viz.},
\begin{equation} \label{eq:Heff}
  \mathcal{H}_{\rm eff} =  -\frac{4G_F}{\sqrt{2}} V_{tb}V_{ts}^\ast
  \frac{\alpha_{\rm em}}{4\pi} \left[C_7 \, \mathcal{O}_7 + C_9 \, \mathcal{O}_9
                                + C_{10} \, \mathcal{O}_{10} \right] + \rm H.c., 
\end{equation}
where $V_{tb}, V_{ts}^\ast$ are the Cabibbo-Kobayashi-Maskawa (CKM)
elements, $G_F$ is the Fermi constant and $\alpha_{\rm em}$ is the fine
structure constant. Denoting the chiral projectors by $P_{L,R} =(1\mp
\gamma_5)/2$, the operators read
\begin{equation}\label{eq:effHamil}
{\mathcal{O}}_{7} = \frac{m_b}{e} 
(\bar{s} \sigma_{\mu \nu} P_R b) F^{\mu \nu} ,\quad\quad 
{\mathcal{O}}_{9} =  
(\bar{s} \gamma_{\mu} P_L b)(\bar{\ell} \gamma^\mu \ell)\, ,\quad\quad
{\mathcal{O}}_{10} =
(\bar{s}  \gamma_{\mu} P_L b)(  \bar{\ell} \gamma^\mu \gamma_5 \ell) \,.
\end{equation}
The Wilson coefficients $C_{7,9,10}$ embody the short-distance
contributions to $\mathcal{O}_i$. Of course, other operators, such as
the standard four-fermion ones ($\mathcal{O}_1 \cdots \mathcal{O}_6$)
as well as the dipole operator $\mathcal{O}_8$ could also
contribute. However, such operators are subdominant and their leading
effects in the decays of interest can be encapsulated in terms of the
changes brought about in $C_{7,9,10}$ through operator mixing (thanks
to renormalization group flows).  As discussed in Appendix
\ref{app:effWCs}, their contributions can be essentially subsumed by
replacing $C_{7,9}$ with $C_{7,9}^{\rm eff}$.
 
 In this paper, we are interested in semileptonic decays of the form
\begin{equation}
  H_1(p_0)\to H_2(p_1)\ell^-(q_1)\ell^+(q_2),
  \label{general_decay}
\end{equation}
where $H_{1,2}$ are hadrons (mesons or baryons as the case may be) and $\ell = e, \mu$. Defining an effective coupling constant
\begin{equation}
  g_{\rm eff} \equiv
  \frac{-4G_F}{\sqrt{2}}V_{tb}V_{ts}^\ast \frac{\alpha_{\rm em}}{4\pi} \ ,
\end{equation} 
and leptonic currents
\begin{equation}
  \mathcal{L}^A_{\mu} = \bar{u}_{\ell_1}(q_1)\Gamma^A_{\mu}v_{\ell_2}(q_2)\ ,
\end{equation}
with $\Gamma^{1,2}_\mu \equiv \gamma_\mu,\gamma_\mu\gamma_5$, 
the matrix element 
for the decay in eqn.(\ref{general_decay}) can be written as
\beq \label{eq:nonrad}
\barr{rcl}
\mathcal{M}_0 &=& \dis \sum_{A=1}^2 \mathcal{L}_\mu ^A \otimes \mathcal{H}^\mu_A,
\\[2ex]
\mathcal{H}^\mu_1(p_0,p_1) &\equiv& \dis
g_{\rm eff} \langle H_2 (p_1)|\bigg\{C_9 \bar{s} \gamma^\mu (1-\gamma_5) b - \frac{2m_b}{q^2}C_7 \bar{s}iq_\nu\sigma^{\mu\nu} (1+\gamma_5) b \bigg\}|H_1 (p_0)\rangle,
\\[2ex]
\mathcal{H}^\mu_2(p_0,p_1) &\equiv& \dis
g_{\rm eff} \langle H_2 (p_1)|C_{10} \bar{s}\gamma^\mu (1-\gamma_5) b  |H_1(p_0)\rangle \ ,
\earr
\eeq
where the long-distance contributions in the $H_1\to H_2$ transition
matrix elements corresponding to different Dirac structures are
parametrized largely in terms of form factors.


\section{QED corrections}\label{sec:qed}
We now proceed to a calculation of the  QED
corrections to the decay width for the process of
eqn.(\ref{general_decay}). This entails the consideration of both
virtual (self-energy as well as vertex) corrections in the effective
theory as also the inclusion of soft emission diagrams.  While it
might be tempting to discount photon lines emanating from neutral
hadrons ($H_{1,2}$), this has to be examined closely, especially in
the context of effective theories (such as the present one), as such
interactions could, presumably, arise as a result of pinching
$n$-point operators. However, whether $H_{1,2}$ are mesons or baryons,
such vertices (whether $H_iH_i\gamma$ or a transition one such as $H_i
H' \gamma$ where $H'$ is an intermediate hadron) have extra powers of
momenta. Consequently, the inclusion of such vertices would not induce
any new infra-red divergences. As for possible new ultra-violet
divergences, of course they could be there, but their effects are
already included in the running down of the Wilson coefficients from
the electroweak scale down to the mass scale of the decaying hadron.


\subsection{Real photon corrections}
\begin{figure}[H]
\vspace*{-2ex}
    \begin{center}
\subfigure[]{%
\begin{tikzpicture}[scale=0.85]
\begin{feynman}
\vertex (a1) at (-0.3,0) {\(H_1(p_0)\)};
\node[blob, minimum size = 0.4 cm] (a4) at (2,0);
\vertex (a2) at (4,-1.5)  {\(\ell^+(q_2)\)};
\vertex (a5) at (4.3,0) {\(H_2(p_1)\)};
\vertex (b1) at (4,1.5) {\(\ell^-(q_1)\)};
\vertex (g1) at (2.5,1.5);
\vertex (g2) at (3.,0.75);
\diagram* {
{[edges=fermion]
(a1) -- [very thick] (a4) -- [very thick] (a5),
(a2) -- (a4) -- (b1),
};
(g1) -- [photon, edge label'=\(\gamma(k)\)] (g2),
};
\end{feynman}
\end{tikzpicture}}
\subfigure[]{%
\begin{tikzpicture}[scale=0.85]
\begin{feynman}
\vertex (a1) at (-0.3,0) {\(H_1(p_0)\)};
\node[blob, minimum size = 0.4 cm] (a4) at (2,0);
\vertex (a2) at (4,-1.5)  {\(\ell^+(q_2)\)};
\vertex (a5) at (4.3,0) {\(H_2(p_1)\)};
\vertex (b1) at (4,1.5) {\(\ell^-(q_1)\)};
\vertex (g1) at (2.5,-1.5);
\vertex (g2) at (3.,-0.75);
\diagram* {
{[edges=fermion]
(a1) -- [very thick] (a4) -- [very thick] (a5),
(a2) -- (a4) -- (b1),
},
(g2) -- [photon, edge label'=\(\gamma(k)\)] (g1),
};
\end{feynman}
\end{tikzpicture}}
\end{center}
\vspace*{-2ex}
\caption{Emission of soft photon from the external lepton pair.\label{fig:realphoton}}
\end{figure}
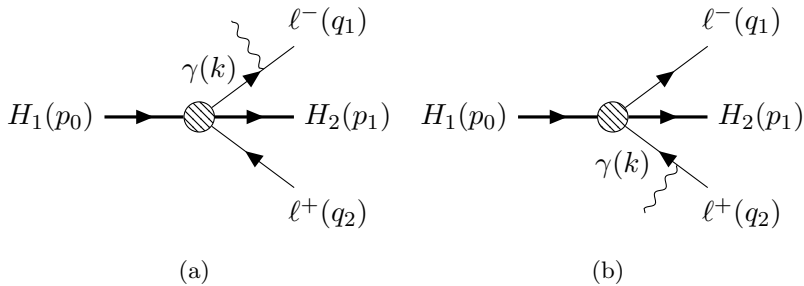

We begin by considering soft real photon emissions associated with the
underlying process of eqn.(\ref{general_decay}) with $H_{1,2}$ being
neutral. As discussed above, in the soft limit (the photon momentum
$k_\mu \to 0$), nonzero contributions would accrue only when such
photons are emitted from the fermion lines. 
The corresponding diagrams, to the leading order, are shown
in Fig.\,\ref{fig:realphoton}.
\noindent
To the lowest order in QED, the amplitude for $H_1(p_0) \to H_2(p_1)
\ell^-(q_1) \ell^+(q_2) \gamma(k)$ takes the form (here $\epsilon(k)$
denotes the photon polarization)
\begin{align} \label{eq:had1}
\mathcal{M}_{\rm real } &= - \bigg[\mathcal{M}_0\sum_{i=1,2}\eta_i Q_i  \frac{q_i.\epsilon^\ast(k)}{q_i.k} +\mathcal{M}_{\rm lept} (k) \bigg]\,,
\end{align}
where $Q_i (i=1,2)$ denote the electric charges of the leptons in
 units of the elementary charge $e$ and
$\eta_i = + (-)$ for outgoing (incoming) particle. The first term on
the r.h.s. of eqn.\eqref{eq:had1} is analogous to Low's soft photon
amplitude consisting of the non-radiative amplitude $\mathcal{M}_0$
given in eqn.\eqref{general_decay}, and a universal function
$q_i\cdot\epsilon^\ast(k) / (q_i\cdot k)$ which is ${\cal O}(k^{-1})$
and, hence, infrared-divergent.  The second term ($\mathcal{M}_{\rm
  lept}$) is finite in both the soft and collinear limits and reads
\begin{align}
 \mathcal{M}_{\rm lept} &= e \mathcal{H}^{\mu}_A(p_0,p_1)\bar{u}_{\ell_1}(q_1) \left\{\gamma^\alpha \frac{\slashed{k}}{2q_1.k}\Gamma_\mu^A  - \Gamma_\mu^A  \frac{\slashed{k}}{2q_2.k}\gamma^\alpha \right\}v_{\ell_2}(q_2)\epsilon^\ast_\alpha(k)\,.
\end{align}
Not only is $\mathcal{M}_{\rm real }$ gauge invariant, but,
individually, so are the Low term as well as $\mathcal{M}_{\rm lept}$.
Squaring $\mathcal{M}_{\rm real}$, summing over the soft photon
polarizations, and integrating over the phase space yields the
differential decay width for $H_1(p_0) \to H_2(p_1) \ell^-(q_1)
\ell^+(q_2) \gamma(k)$. To $\mathcal{O}(\alpha_{\rm em})$, this reads
\begin{eqnarray}\label{eq:decayrad}
\frac{d\Gamma_{\rm real}}{dq^2} = \frac{d\Gamma_0}{dq^2} \times \alpha_{\rm em} B_{\rm real} + \frac{d \Gamma^\prime}{dq^2}\,,
\end{eqnarray}
where $d\Gamma_0/dq^2$ refers to the differential decay rate for the
lowest-order process (\eqref{general_decay}). The first term in
eqn.\eqref{eq:decayrad} arises purely from the squaring of the Low term,
while $d\Gamma^\prime$ encapsulates the rest ({\em viz.}, the square
of $\mathcal{M}_{\rm lept}$ as well as the interference with the Low
term). While all terms in eqn.\eqref{eq:decayrad} are, understandably,
$\alpha_{\rm em}$ suppressed with respect to the lowest-order process
(\eqref{general_decay}), the first term would turn out to be enhanced
by large logarithms, in contrast to ${d\Gamma'}/{dq^2}$. Owing
  to the absence of such enhancements, $ {d\Gamma^\prime}/{dq^2}$ would turn out to be numerically insignificant and
may be neglected.  The multiplicative factor $ B_{\rm real}$ is given
by
\begin{eqnarray}
B_{\rm real} &= &\frac{1}{2 \pi^2}\int \frac{d^3\vec{k} }{2k^0}\Big |\sum_{i=1,2} \sum _{\lambda =1, 2} Q_i \eta_i \frac{ q_i.\epsilon_\lambda^\ast(k)}{q_i.k}\Big | ^2 \,,
\end{eqnarray}
and includes infrared divergences which would be
cancelled by corresponding divergences in the self-energy and vertex
corrections. Given that the source of such divergences are the
$U(1)_{\rm em}$ interactions, these are most easily handled by
introducing a non-zero photon mass $\lambda$ (divergences would
appear as $\lambda \to 0$). Adopting this, we have
\begin{eqnarray}\label{eq:intreal}
B_{\rm real} &=& \frac{-1}{4 \pi^2}\int_{|\vec{k}|<{\Delta E_s}}
\frac{d^3\vec{k}}{\sqrt{\vec{k}^2 + \lambda^2}} \bigg(
\frac{q_1^\alpha}{q_1.k}- \frac{q_2^\alpha}{q_2.k}\bigg)^2\,
= 
\int_{|\vec{k}|<{\Delta E_s}}
\frac{d|\vec{k}|}{\sqrt{\vec{k}^2+\lambda^2}}\times \mathcal{A}\, ,
\end{eqnarray}
where, for later convenience, we define $\mathcal{A}$ as
\begin{eqnarray}
\mathcal{A} \equiv \frac{-|\vec{k}|^2}{4\pi^2}\int d\Omega \bigg( \frac{q_1^\alpha}{q_1.k}- \frac{q_2^\alpha}{q_2.k}\bigg)^2\ . 
\end{eqnarray}
Physically, the term $\alpha_{\rm em}\mathcal{A}$ is a measure of the
probability of photon emission and rises logarithmically with the
energy of externally charged leptons. The introduction of the photon
momentum cutoff $\Delta E_s$ needs a comment. The photon-emission
process under discussion is distinguishable from the lowest order
process of eqn.\eqref{general_decay} only if the photon is detectable;
this requires that the photon has a minimum energy $\Delta E_s$ and,
furthermore, is separable, in the detector, from the outgoing
particles ({\em i.e.}, the angle between the corresponding momenta is
larger than the detector resolution). Clearly, our results (at least
the interim ones) would depend on both $\lambda$ and $\Delta
E_s$. Here, $\Delta E_s$ denotes the infrared photon-energy cutoff in the dilepton rest frame (see Appendix~\ref{app:Breal}), introduced to separate soft-photon emission from experimentally resolvable radiation. It is a theoretical resolution parameter and does not correspond to an explicit photon-energy veto imposed in the experimental analyses. 

The integral $\mathcal{A}$ is a function of the fermion velocity
$\beta \equiv \sqrt{1- 4 m^2/q^2}$ with $m$ denoting the mass of
  the outgoing leptons. Defining
  \begin{equation}
\lb \equiv \ln \frac{1-\beta}{1+\beta} \ ,
\end{equation}
  we have
\begin{eqnarray}
  \mathcal{A} = \frac{-2}{\pi}\left(1+\frac{1+\beta^2}{2 \beta} \lb \right).
  \label{A-defn}
\end{eqnarray}
Using the analytical results for $B_{\rm real}$ (as presented in
Appendix \ref{app:Breal}), the differential decay rate (with the
photon momentum restricted to $|\vec k|
  < \Delta E_s$) is, then, given by
\begin{eqnarray}\label{eq:finalrad}
  \frac{d\Gamma_{\rm real}}{d q^2} &= & \frac{d\Gamma_0}{d q^2} \;
  \frac{\alpha_{\rm em}}{\pi} \;
  \Bigg\{ \left[ 1+ \left( \frac{1+\beta^2}{2\beta} \right)
    \lb \right]
  \left( \ln\frac{\lambda^2}{m^2} 
          - \ln\frac{4 \Delta E_s^2}{m^2}  \right)
\nn\ \\
& & \hspace*{4em}- \;
    \frac{1+\beta^2}{2\beta} \left[ 2 \rm Li_2 \left( \frac{2\beta}{1+\beta}\right)+ \frac{1}{2}\lb^2 \right]
 - \,     \frac{1}{\beta}\lb\Bigg\} \,.
\end{eqnarray}
Several infinities are contained in eqn.\eqref{eq:finalrad}. The term
proportional to $\ln \left(\lambda^2/m^2\right)$ is a spurious one; an
intermediate stage artefact of the regularization prescription, it
would cancel against similar contributions from the virtual
corrections (to the self-energies and the vertex).  The more
``physical'' singularities can be broadly classified into collinear,
soft and soft-collinear types.  The terms in eqn.\eqref{eq:intreal} going
as $q_i^2 / (q_i \cdot k)^2$ diverge as the photon becomes very soft
($\Delta E_s \to 0$) or the lepton mass vanishes.  The ensuing (soft)
divergences are manifested, in eqn.\eqref{eq:finalrad}, by the term
proportional to $\ln\left(4 \Delta E_s^2/m^2 \right)$.  The remaining
terms in eqn.\eqref{eq:finalrad} diverge as $m \to 0$ (since, then,
$\beta \to 1$), and receive contributions from the term in
eqn.\eqref{eq:intreal} proportional to $(q_1 \cdot q_2) / ((q_1 \cdot k) \;
(q_2 \cdot k))$. It should be noted that the size of the (finite piece
of the) correction is larger for a smaller lepton mass. This is
understandable as a lighter particle radiates more profusely, and this
is a feature that would persist even for the virtual corrections.

\subsection{One-loop virtual corrections}
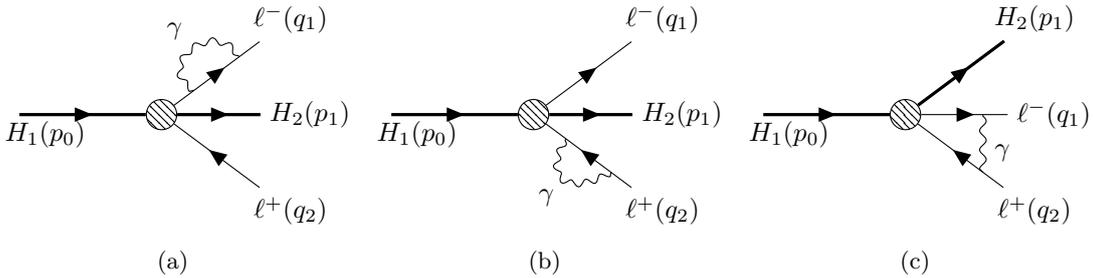
\begin{figure}[H]
    \begin{center}
  {\small
    \subfigure[]{%
\begin{tikzpicture}[scale=0.85]
\begin{feynman}
  \vertex (a1) at (-0.2,0) ;
  \vertex (a0) at (0.2,-0.3) {\(H_1(p_0)\)};
\node[blob, minimum size = 0.4 cm] (a4) at (2,0);
\vertex (a2) at (4,-1.5)  {\(\ell^+(q_2)\)};
\vertex (a5) at (4.3,0) {\(H_2(p_1)\)};
\vertex (b1) at (4,1.5) {\(\ell^-(q_1)\)};
\vertex (g1) at (2.5,0.38);
\vertex (g2) at (3.2,0.9);
\diagram* {
{[edges=fermion]
(a1) -- [very thick] (a4) -- [very thick] (a5),
(a2) -- (a4) -- (b1),
},
(g2) -- [photon, out = 120, in = 150, looseness = 2.2, edge label'=\(\gamma\)] (g1),
};
\end{feynman}
\end{tikzpicture}}
\subfigure[]{%
\begin{tikzpicture}[scale=0.85]
\begin{feynman}
\vertex (a1) at (-0.2,0);
\vertex (a0) at (0.2,-0.3) {\(H_1(p_0)\)};
\node[blob, minimum size = 0.4 cm] (a4) at (2,0);
\vertex (a2) at (4,-1.5)  {\(\ell^+(q_2)\)};
\vertex (a5) at (4.3,0) {\(H_2(p_1)\)};
\vertex (b1) at (4,1.5) {\(\ell^-(q_1)\)};
\vertex (g1) at (2.5,-0.38);
\vertex (g2) at (3.2,-0.9);
\diagram* {
{[edges=fermion]
(a1) -- [very thick] (a4) -- [very thick] (a5),
(a2) -- (a4) -- (b1),
},
(g1) -- [boson, out = -120, in = -150, looseness = 2.2,  edge label'=\(\gamma\)] (g2),
};
\end{feynman}
\end{tikzpicture}}
\subfigure[]{%
\begin{tikzpicture}[scale=0.85]
\begin{feynman}
\vertex (a1) at (-0.2,0);
\vertex (a0) at (0.2,-0.3) {\(H_1(p_0)\)};
\node[blob, minimum size = 0.4 cm] (a4) at (2,0);
\node at (3.5,-0.6) {$\gamma$};
\vertex (a2) at (4,-1.5) {\(\ell^+(q_2)\)};
\vertex (a5) at (4.3,0) {\(\ell^-(q_1)\)};
\vertex (b1) at (4,1.5) {\(H_2(p_1)\)}; 
\vertex (g1) at (3.2,0);
\vertex (g2) at (3.2,-0.9);
\diagram* {
{[edges=fermion]
(a1) -- [very thick] (a4) --  (a5),
(a2) -- (a4) -- [very thick] (b1),
},
(g1) -- [boson, out = 0, in = 0, looseness = 0] (g2),
};
\end{feynman}
\end{tikzpicture}}
  }
\end{center}
\caption{One-loop virtual correction diagrams. The first two
  diagrams lead to self energy corrections for the external leptons, 
  whereas the last one corresponds to vertex
  correction.}
\label{fig:vertex}
\end{figure}

The QED virtual corrections are of two types, namely the
  self-energy renormalizations of the charged leptons and the
  correction to the four-point vertex. The one-loop diagrams in the effective theory are displayed in Fig.\,\ref{fig:vertex}. 
The wavefunction renormalizations ($Z_\ell \equiv 1 + \delta Z_\ell$), 
  understandably, appear only as multiplicative factors acting on the external legs without changing the structure of the matrix element. Thus, to ${\cal O}(\alpha_{\rm em})$, the resultant additional contribution can be expressed as
\begin{eqnarray}\label{eq:selfformula}
\mathcal{M}_{\rm self} =\frac{ \mathcal{M}_0 }{2}\left(\delta Z_{\ell^-} + \delta Z_{\ell^+}\right)\,,
\end{eqnarray}
where 
\beq\label{eq:self}
\barr{rcl}
\delta Z_{\ell^-} &=& \dis
\frac{\alpha_{\rm em}}{4\pi}\left[2- B_0(q_1^2, 0, m^2) + 4m^2 \frac{\partial}{\partial q_1^2} B_0 (q_1^2, \lambda^2, m^2)\right]\,, \\[2ex]
\delta Z_{\ell^+} &=& \dis \delta Z_{\ell^-} (q_1^2 \to q_2^2) \ ,
\earr
\eeq
with the Passarino-Veltman functions $B_0$, or their derivatives,
being evaluated for on-shell leptons, {\em i.e.}, at $q_i^2 = m^2$.
 Thus, $\delta Z_{\ell^+} = \delta
Z_{\ell^-} \equiv \delta Z_{\ell}$. As before, we have introduced a
mass $\lambda$ for the photon (as an IR regulator). It should be noted
that the two-point functions $B_0$ are associated with UV divergences
but are free from divergences in the IR. On the other hand, for the
derivatives, the situation is the opposite.

Coming to the vertex correction ({\em i.e.}, the last diagram in Fig.\,\ref{fig:vertex}), the corresponding amplitude can be written as
\begin{eqnarray}\label{eq:vertex}
\mathcal{M}_{\rm vertex } =H_\mu^A(p_0,p_1)\int \frac{d^4 k}{(2\pi)^4} \frac{B_A^\mu(q_1,q_2,k,m,m)}{\big(k^2 - \lambda^2\big)\big((q_1 +k)^2 - m^2\big)\big(( q_2-k)^2 - m^2\big)},
\end{eqnarray}
where the (pseudo)-vectorial current is given by
\begin{equation}
  \begin{array}{rcl}
    B_A^\mu(q_1,q_2,k,m,m)
    &\equiv & \displaystyle
    i e^2 \bar{u}_{\ell_1}(q_1) \gamma^\alpha(\slashed{q_1} + \slashed{k} +m)\Gamma_A^\mu (\slashed{q_2} - \slashed{k} - m)\gamma_\alpha  v_{\ell_2}(q_2)
    \\[0.ex]
    &= & \displaystyle
    i e^2 \bar{u}_{\ell_1}(q_1)
    \left( 2 q_1^\alpha + \gamma^\alpha \slashed{k}\right)
    \Gamma_A^\mu \left(2 q_{2\alpha} - \slashed{k} \gamma_\alpha\right)
    v_{\ell_2}(q_2),
  \end{array}
  \label{eq:vertexampl}
\end{equation}
with $\Gamma_A^\mu = \gamma^\mu, \gamma^\mu \gamma_5$ for $A = 1,
2$. In reaching the second line, the equations of motion have been
used. And while it might naively seem that $B^\mu_A$ is independent of
the mass $m$, it is easy to see that the terms linear in $k$ would, on
the equations of motion being applied, lead to terms proportional to
$m$.  It is also easy to see that the $ 4 q_1\cdot q_2 \Gamma_A^\mu$
term, on integration, is free of UV divergence, but does suffer from
IR divergences. Finally, while this term has a structure exactly
analogous to the tree-level term, the same is not true of the rest and
these would, typically, lead to terms that are not simply
factorizable.  Nonetheless, on performing the momentum integration and
some simple Dirac algebra, the amplitude takes a simple form, namely
\begin{equation}\label{eq:vertexsplit}
  \mathcal{M}_{\rm vertex} =  \frac{\alpha_{\rm em}}{ \pi}
  \left[ - (q_1.q_2) C_0 \mathcal{M}^0 + \left\{\frac{1}{2} g_{\alpha\beta}C^{\alpha\beta}+(q_{1\alpha} -q_{2\alpha} )C^\alpha \right\}\mathcal{M}^0  + \mathcal{M}^\prime
    \right].
\end{equation}
with the non-factorizable term being 
\begin{equation}
  \mathcal{M}^\prime = 
    \Big[m  C^\mu \bar{u}_{\ell_1}(q_1) v_{\ell_2}(q_2) - \big(C^{\alpha\mu} + ( q_1^\mu - q_2^\mu)C^{\alpha}\big) \bar{u}_{\ell_1}(q_1)\gamma_\alpha v_{\ell_2}(q_2)\Big]
    \mathcal{H}_{\rm 1\mu}(p_0,p_1) ,
\end{equation}
for the vector-current and
\begin{equation}
\begin{array}{rcl}
  \mathcal{M}^\prime & =&
        \displaystyle
            \Big[
m  C^\mu \bar{u}_{\ell_1}(q_1)\gamma_5 v_{\ell_2}(q_2)
 - \big(C^{\alpha\mu} + ( q_1^\mu - q_2^\mu)C^{\alpha}\big) \bar{u}_{\ell_1}(q_1)\gamma_\alpha \gamma_5 v_{\ell_2}(q_2) \\
&& \displaystyle \hspace{1ex} - m  C_{\alpha} \bar{u}_{\ell_1}(q_1)\gamma^\alpha\gamma^\mu\gamma_5 v_{\ell_2}(q_2)\Big]\mathcal{H}_{\rm 2\mu}(p_0,p_1) \ ,
\end{array}
\end{equation}
for the pseudovector current.
Here the hadronic matrix elements $\mathcal{H}^\mu_{\rm 1,2}(p_0,p_1)$ are as
given in eqn.\eqref{eq:nonrad}, and
all the Passarino-Veltman
three point functions (whether scalar $C_0$, vector $C_{\mu}$ or
tensor $C_{\mu\alpha}$) are evaluated with an identical set of
arguments, {\em viz.}, $C_i\equiv C_i(q_1,-q_2,\lambda,m,m)$, and
the explicit expression of $C_0$ is
given in Appendix \ref{app:veltman}. Once again, note that, in the UV,
the scalar and vector functions $C_{0,\mu}$ are convergent, while the
tensor functions $C_{\mu\alpha}$ are logarithmically divergent owing
to the presence of a term quadratic in $k$. For the same reason, this
term is free of an IR-divergence.  In contrast, $C_0$ is IR-divergent
while being UV-convergent. The preceding arguments lead to the
important conclusion that the second and third terms of
eqn.(\ref{eq:vertexsplit}) are free of infrared divergences and vanish
in the soft photon limit.  Consequently, post-renormalization,
these are of little
importance in the remainder of the analysis, and could, in effect, be
neglected altogether. Neglecting this would have the added
advantage that the one-loop vertex correction would now be entirely
factorizable in terms of the tree-level amplitude, thereby considerably
simplifying the calculations.
%

\subsection{Total $\mathcal{O}(\alpha_{\rm em})$ QED correction}
To the leading order, then, the modulus-squared of the exclusive
  three-body decay matrix element is given by
\beq \label{eq:totalmod}
\barr{rcl}
|\mathcal{M}|^2 & = & \dis \left| \mathcal{M}_0 + \mathcal{M}_{\rm vertex}
+ \mathcal{M}_{\rm self} \right|^2
\\[1ex]
& = & \dis 
|\mathcal{M}_0|^2 + 2 \re \big( \mathcal{M}_0 \mathcal{M}_{\rm vertex}^\ast\big)  + 2 \re \big( \mathcal{M}_0 \mathcal{M}_{\rm self}^\ast  \big) + \mathcal{O}(\alpha_{\rm em}^2)\,,
\earr
\eeq
with the second and third terms in the second line including the
  divergences. Effecting the integration over the phase space and the
  usual summing and averaging over spins, the decay
rate for the 3-body final state process, {\em viz.},
$H_1(p_0) \to H_2(p_1) \ell^- (q_1) \ell^+ (q_2) $, to $\mathcal{O}(\alpha_{\rm em})$, is
%
\beq \label{eq:treevirtual}
d\Gamma_{\rm 3-body} = d\Gamma_0 \Big\{1 - \frac{2\alpha_{\rm em}}{ \pi} (q_1.q_2) \re \left( C_0\right) + 2\delta Z_{\ell}  \Big\}\,.
\eeq
To this, must be added the rate for the four-body decay (including a
soft/collinear photon radiation), and the usual UV-renormalization
effected. The resultant, {\em viz.}
\begin{eqnarray}\label{eq:total}
  d \Gamma^{\rm tot} = 
  d \Gamma_{\rm 3-body} + d \Gamma_{\rm 4-body}\, ,
\end{eqnarray}
is free of infrared divergences as these are cancelled between the two
pieces. Using eqns. \eqref{eq:finalrad}
and \eqref{eq:treevirtual} in eqn.\eqref{eq:total}, the total differential decay rate, upto
$\mathcal{O}(\alpha_{\rm em})$, can be written as
\begin{eqnarray}\label{eq:qedcorr}
\frac{d \Gamma^{\rm tot}}{d q^2}  
&=& \frac{d \Gamma_0}{d q^2}\left(1 + \delta^\ell\right),
\end{eqnarray}
where ${d \Gamma_0}/{d q^2}$ is the differential decay rate for the
(parent) three-body decay process, and the correction factor
$\delta^\ell$ for the lepton $\ell$ is
\begin{equation}\label{eq:delta}
\begin{array}{rcl}
  \delta^\ell &=& \displaystyle
  \frac{-\alpha_{\rm em}}{\pi} \Bigg\{ \left[ 1+ \left( \frac{1+\beta^2}{2\beta} \right)\lb \right] \ln\left(\frac{4 \Delta E_s^2}{m^2} \right)
  +  \frac{1}{\beta} \lb
  \\
  && \displaystyle \hspace*{3em}
  + \left(\frac{1+\beta^2}{2\beta}\right)\left[ 4 \rm Li_2 \left(\frac{2\beta}{1+\beta}\right) + \lb^2 - \pi^2 
 \right] + \frac{1}{2}\left(2+B_0\right)  \Bigg\} .
\end{array}
\end{equation}
The two point scalar function $B_0\left(m^2,0,m^2\right)$ in the above
originates from the self energy corrections (post regularization) of
the lepton legs and is free from infrared
divergences.

The logarithamic nature of $\delta^\ell$ is reminiscent of the
well-known radiative tail. Not only is the correction factor a
negative quantity (implying that QED corrections reduce the partial
widths), it is numerically larger for smaller masses of the lepton,
which is but a manifestation of the fact that lighter particles
radiate more.  Consequently, ratios of branching fractions, such as
$R_K$, stand to receive nontrivial corrections owing to such
radiations.

Before we actually compare our results to observables, it needs to be
realized that what we have calculated above is only the first
nontrivial order. Calculating to higher loops, however, is a very
difficult task especially in effective theories such as we have
here. What is more tractable, though, is to account for the fact that
the charged leptons would, in general, radiate an indefinite number of
photons. Fortunately enough, full higher order corrections need not be
computed, thanks to the well-known result~\cite{Yennie:1961ad} that
IR-divergence cancellations proceed in higher orders just as they do in
in the lowest order. This allows us to 
simply exponentiate the lowest-order
corrections. This exponentiation (resummation of the leading logarithms)
  leads to a final result containing double logarithmic
  terms, \emph{viz.,}

\begin{equation}\label{eq:correxp}
\begin{array}{rcl}
  \displaystyle \frac{d\Gamma^{\rm tot}_{\rm resum}}{d q^2} &=&
  \displaystyle \frac{d\Gamma_0}{d q^2} F (\alpha_{\rm em} \mathcal{A})
  \\[0.5ex]
  & \times & \displaystyle 
  \exp\left[\frac{-\alpha_{\rm em}}{\pi}
    \left\{ 1+ \frac{1+\beta^2}{2\beta} \, \lb
    \right\} \left\{ \ln \frac{4 \Delta E_s^2}{m^2} + \lb
      \right\}  \right ]
  \\[1.5ex]
  &\times & \displaystyle \left [ 1- \frac{\alpha_{\rm em}}{\pi}
    \left( \frac{1+\beta^2}{2\beta}  \left\{
    4\rm  Li_2 \left( \frac{2\beta}{1+\beta} \right)- \pi^2 \right\}
    + \frac{1-\beta}{\beta} \lb
      +\frac{2+B_0}{2} \right)\right] \\[1.5ex]
  &\equiv& \displaystyle
  \frac{d\Gamma_0}{d q^2} \left( 1+ \delta^\ell_{\rm resum} \right) \ ,
\end{array}
\end{equation}
with $\mathcal{A}$ being given by eqn.(\ref{A-defn}).
The physical requirement that the sum of all soft-photon
energies be less than $\Delta E_s$ leads to the normalization factor
\begin{eqnarray}\label{eq:falphaA}
F(\alpha_{\rm em} \mathcal{A})&=& \frac{e^{-\alpha_{\rm em}\mathcal{A} C}}{\Gamma(1 + \alpha_{\rm em} \mathcal{A})} =  1- \frac{\pi^2(\alpha_{\rm em} \mathcal{A})^2}{12} +...
\end{eqnarray}
where $C$ is the Euler constant.
To a good approximation, then, $F(\alpha_{\rm em} \mathcal{A}) \approx 1$.

The factor $\delta^{\ell}_{\rm resum}$ encapsulates the
next-to-leading-logarithm (NLL) corrections to the process under
consideration and is, hence, expected to be a better estimate of the
corrections pending full NNLO calculations.  It is worth noting that
the absolute value of $\delta^{\ell}_{\rm resum}$, for a fixed $\Delta
E_s$, is a little less than that of $\delta^{\ell}$.  This departure
of $\delta^\ell_{\rm resum}$ from $\delta^\ell$ is greater for
smaller lepton masses and smaller values of $\Delta E_s$, serving to
bring the theoretical expectations for the electron channel closer to
the LO calculations (than is the case for the NLO
numbers).
%

\subsection{Numerical Results}\label{sec:num}
Having effected analytical calculations, we now proceed to numerical
evaluations of the corrections. In this, we would not only calculate
the NLO (NLL) correction factors $\delta^\ell$($\delta^\ell_{\rm
  resum}$) for each of electronic and muonic final states as functions
of the leptonic invariant mass ($q^2$), but also their effect on
\emph{LFU observables} such as ratios of the differential decay widths.

The QED-NLO ($\delta^\ell$) and NLL ($\delta^\ell_{\rm resum}$) corrections are
displayed in Fig.\,\ref{fig:deltaq2}. Several features are immediately obvious:

\begin{figure}[ht!]
\begin{center}
 \includegraphics[scale=0.325]{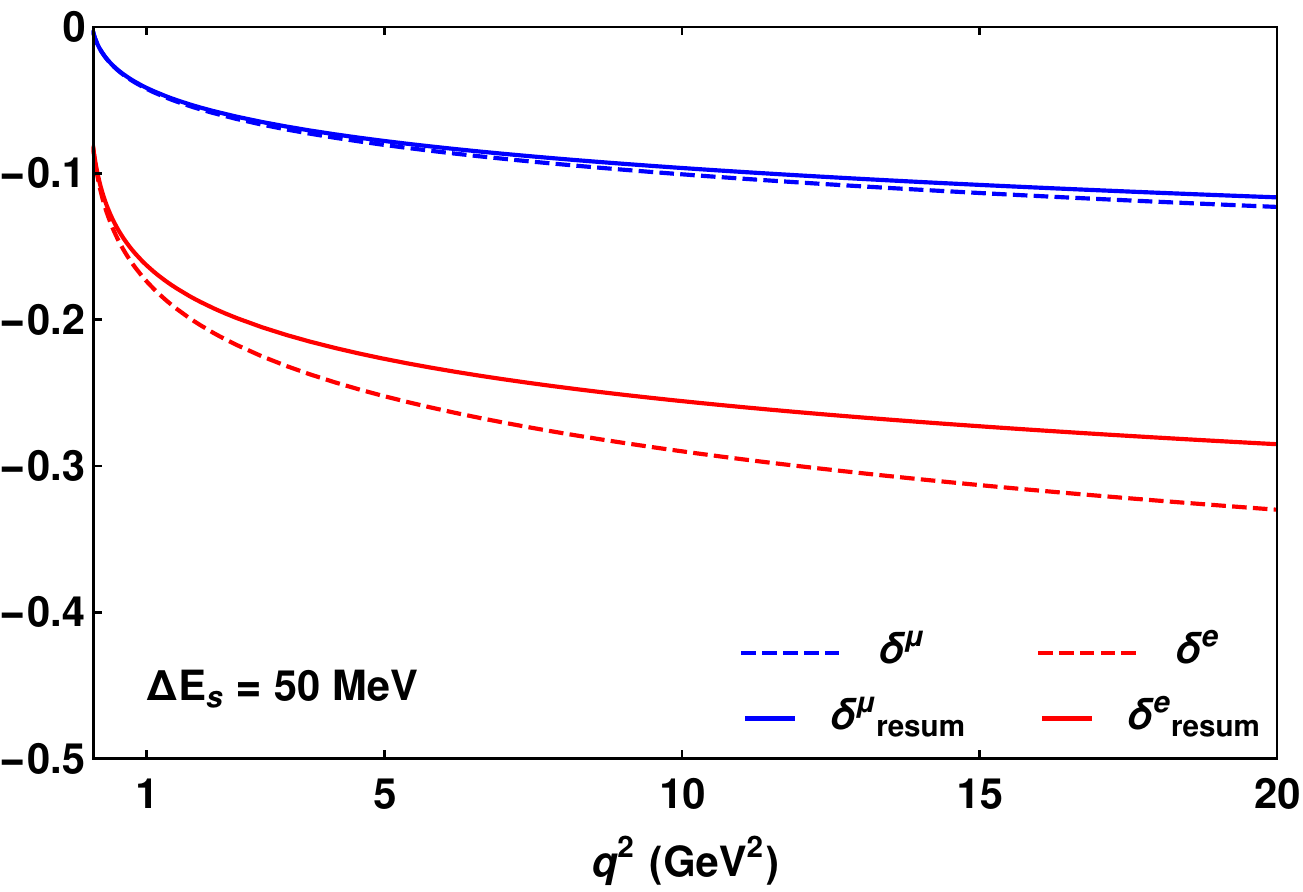}
 \includegraphics[scale=0.325]{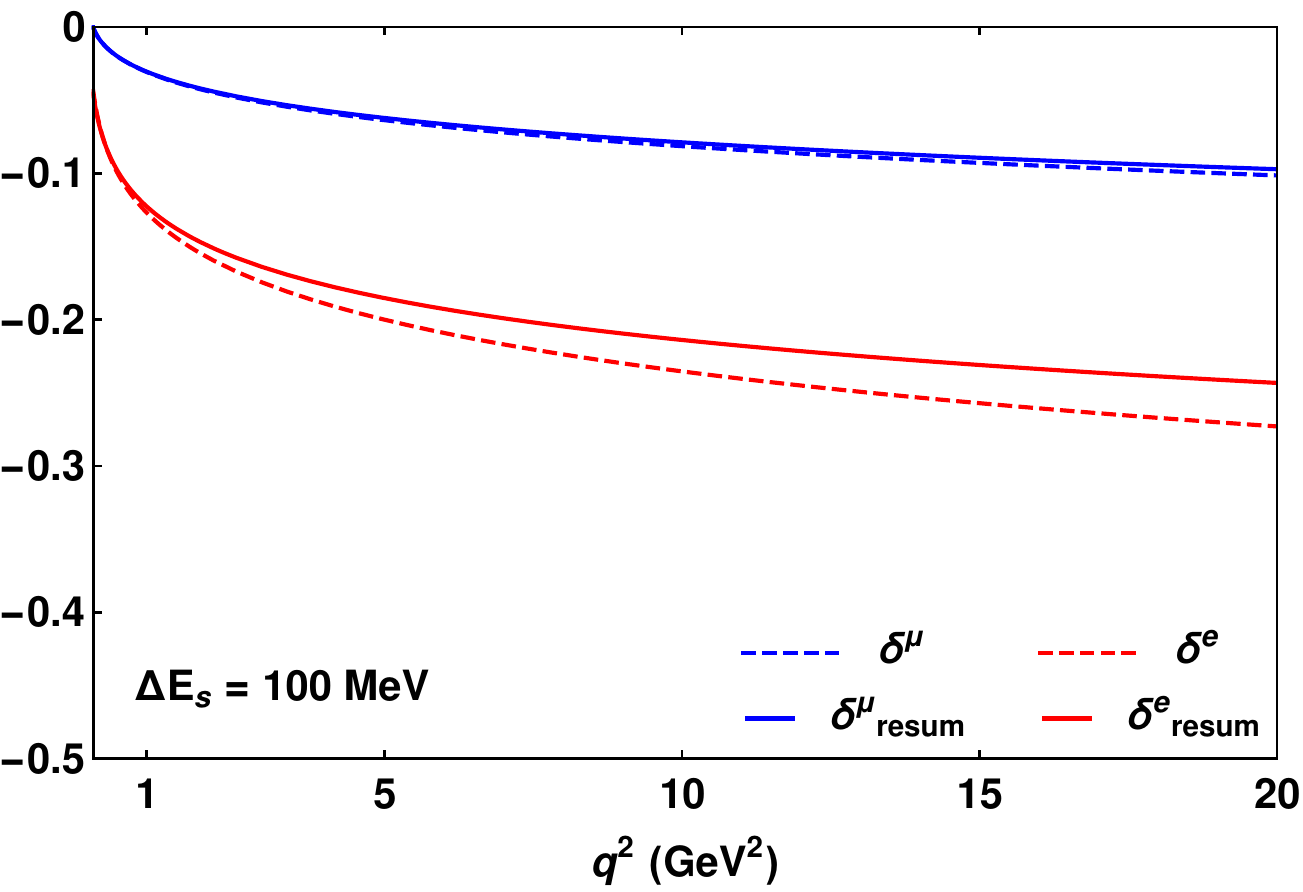}
\end{center}
\caption{QED correction factors $\delta^\ell$ and $\delta^\ell_{\rm
     resum}$ where $\ell= e,\mu $ as a function of $q^2$ for $\Delta E_s
   =$ 50 MeV (left panel) and $\Delta E_s =$ 100 MeV (right
   panel). }
 \label{fig:deltaq2}
\end{figure}

\begin{figure}[ht!]
\begin{center}
\includegraphics[scale=0.325]{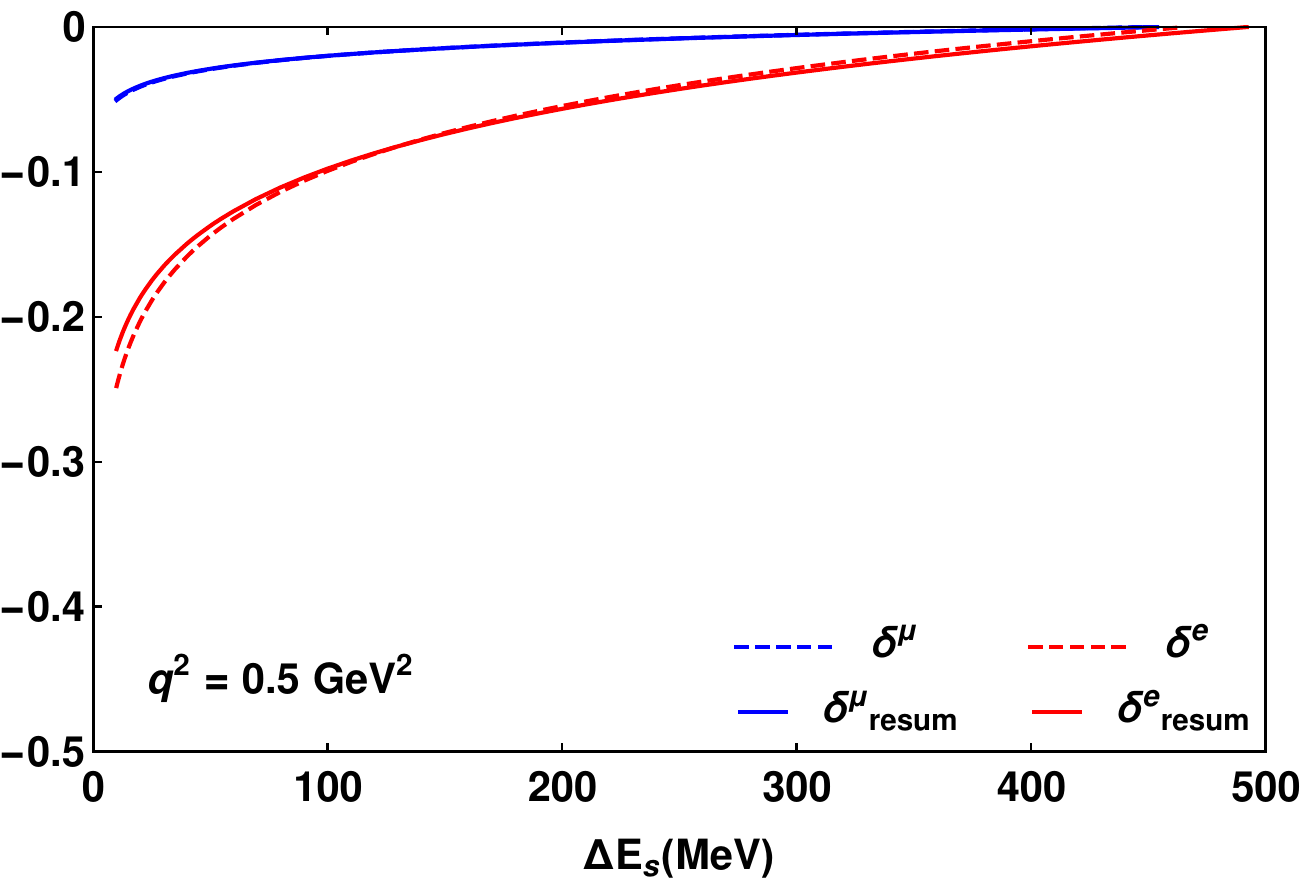}
\includegraphics[scale=0.325]{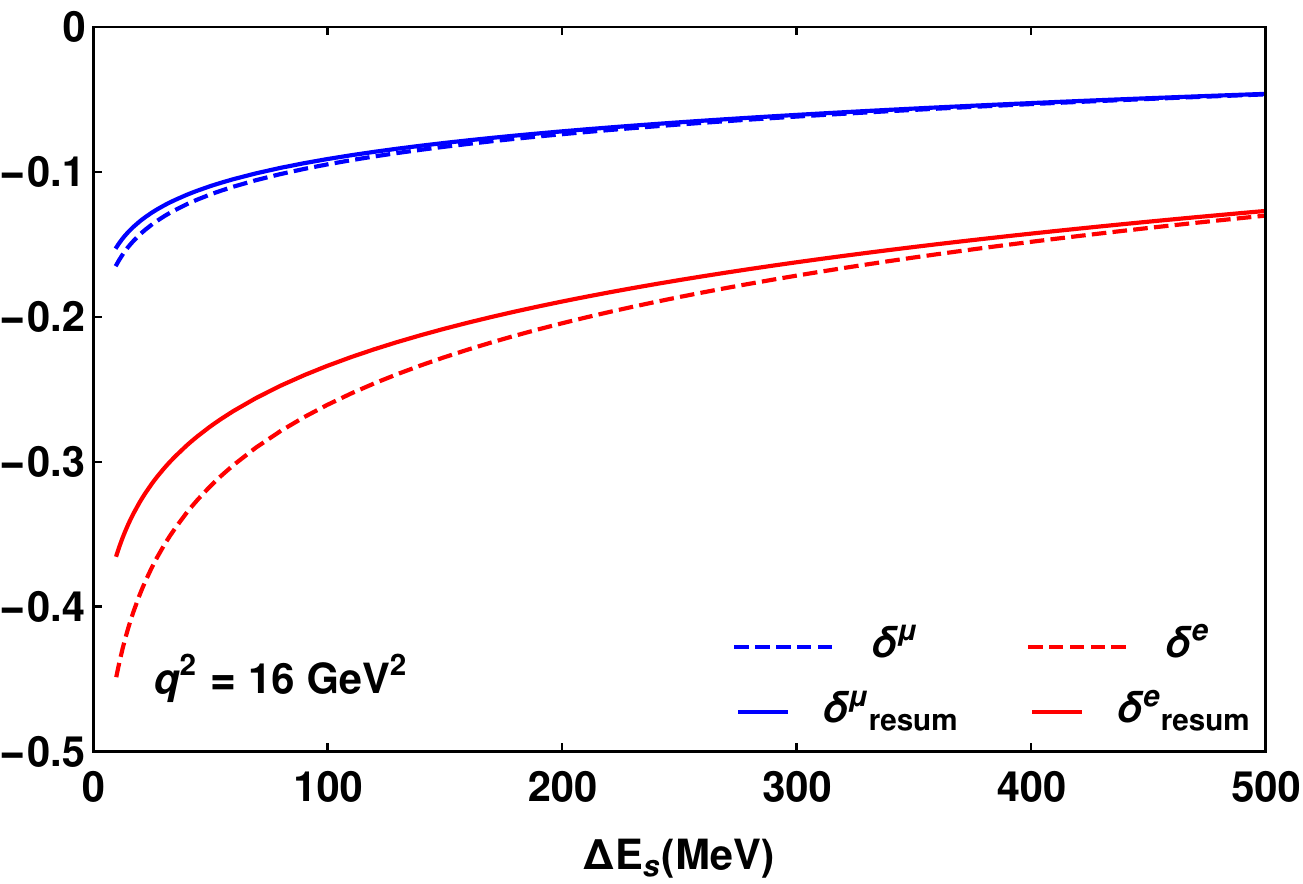}
\caption{QED correction factors $\delta^\ell$ and $\delta^\ell_{\rm
    resum}$ where $\ell = e,\mu$ with respect to soft-photon momentum cutoff $\Delta E_s$ for fixed $q^2$ = \{0.5$\rm GeV^2$, 16 $\rm
  GeV^2$\}.\label{fig:deltakmax}} 
\end{center}

\end{figure}

\begin{itemize}
\item The corrections are always negative. In other words, the decay widths
  are smaller as compared to the tree-order values.
\item Expectedly, the corrections grow with $q^2$, the invariant mass
  of the lepton pair.
\item That the corrections are much larger for the electron than for the
  muon is understandable. Even classically, the radiation rate is higher
  for a lighter particle, and this feature is carried over to the quantum
  mechanical case.
\item More nontrivially, $|\delta^\ell_{\rm resum}| <
  |\delta^\ell|$. This can be traced to the fact that the act
    of resumming partly ameliorates the large logarithms by
  exponentiating them.
\item Indeed, the size of $|\delta^\ell|$ is uncomfortably large
  for large $q^2$, a consequence of the aforementioned large logarithms (as, for example, in $\lb$). The NLL correction term $\delta^\ell_{\rm resum}$, on the other hand, is small enough for a perturbative treatment to be valid.

\item The amelioration is more pronounced for the electron than for
  the muon. Once again, this can be traced to the fact that the
  aforementioned large logarithms are larger for the electrons on
  account of their smaller masses.
\end{itemize}
The results above have been derived for a given value of the two
cutoffs, namely, the infrared one ($\Delta E_s$)
on the energy of the
photon as well as the ultraviolet cutoff $\Lambda$. We now examine
these dependences in turn. As for the former, a comparison of the left and right
panels of Fig.\,\ref{fig:deltaq2} already tells us that increasing
$\Delta E_s$ reduces $\delta^\ell$ (or $\delta^\ell_{\rm resum}$).
While this might seem counterintuitive given the definition of, say,
$B_{\rm real}$ (see eqn.\eqref{eq:intreal}), note that the latter has
only a logarithmic dependence on $\Delta E_s$. And since $\Delta E_s$
is much smaller than the other scales in the process, increasing it
obviously reduces the magnitude of $B_{\rm real}$ (and, similarly,
each of $\delta^\ell$ and $\delta^\ell_{\rm resum}$).  This behaviour
is valid for a wide range of $\Delta E_s$ (see
Fig.\,\ref{fig:deltakmax}), even beyond the range that one would
nominally use.

\begin{figure}[ht!]
\begin{center}
\includegraphics[scale=0.325]{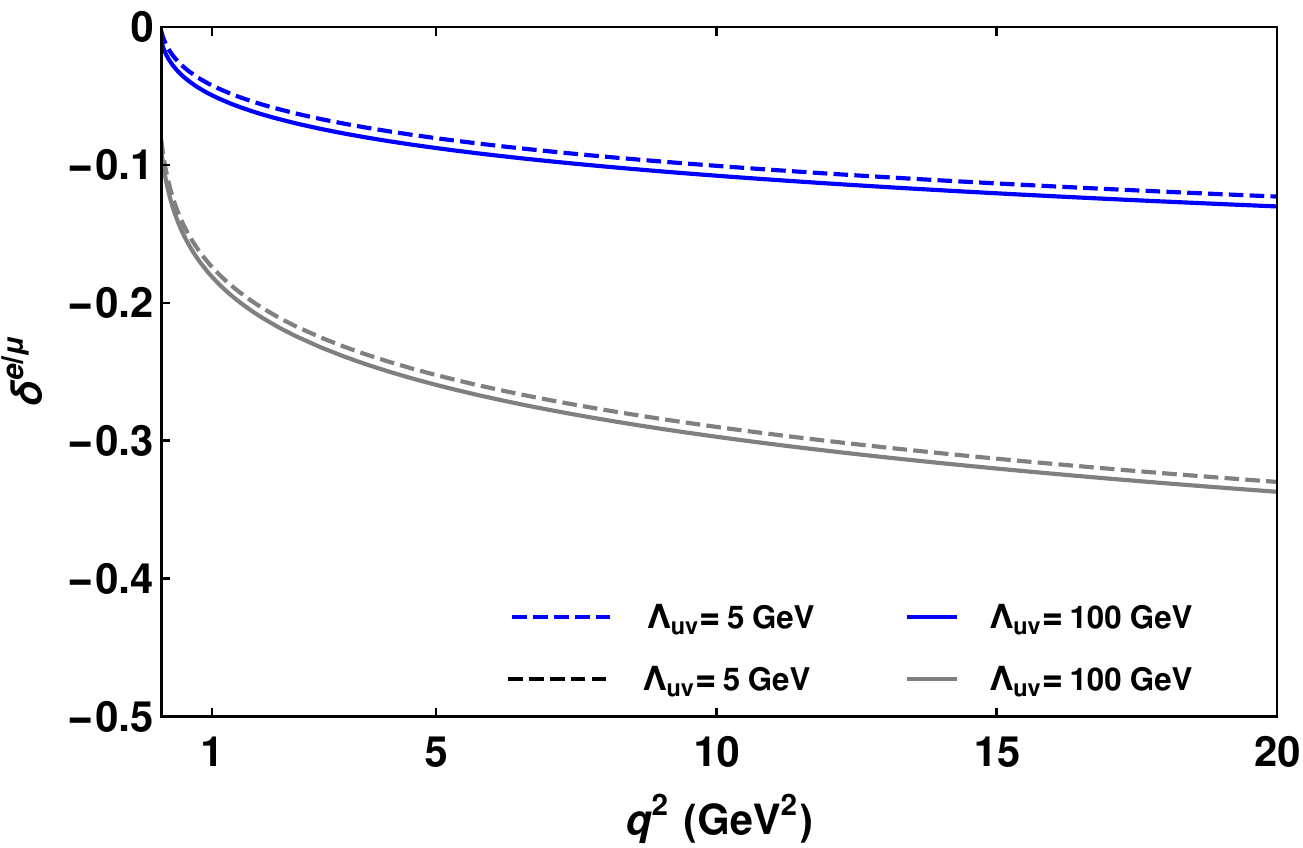}
\includegraphics[scale=0.325]{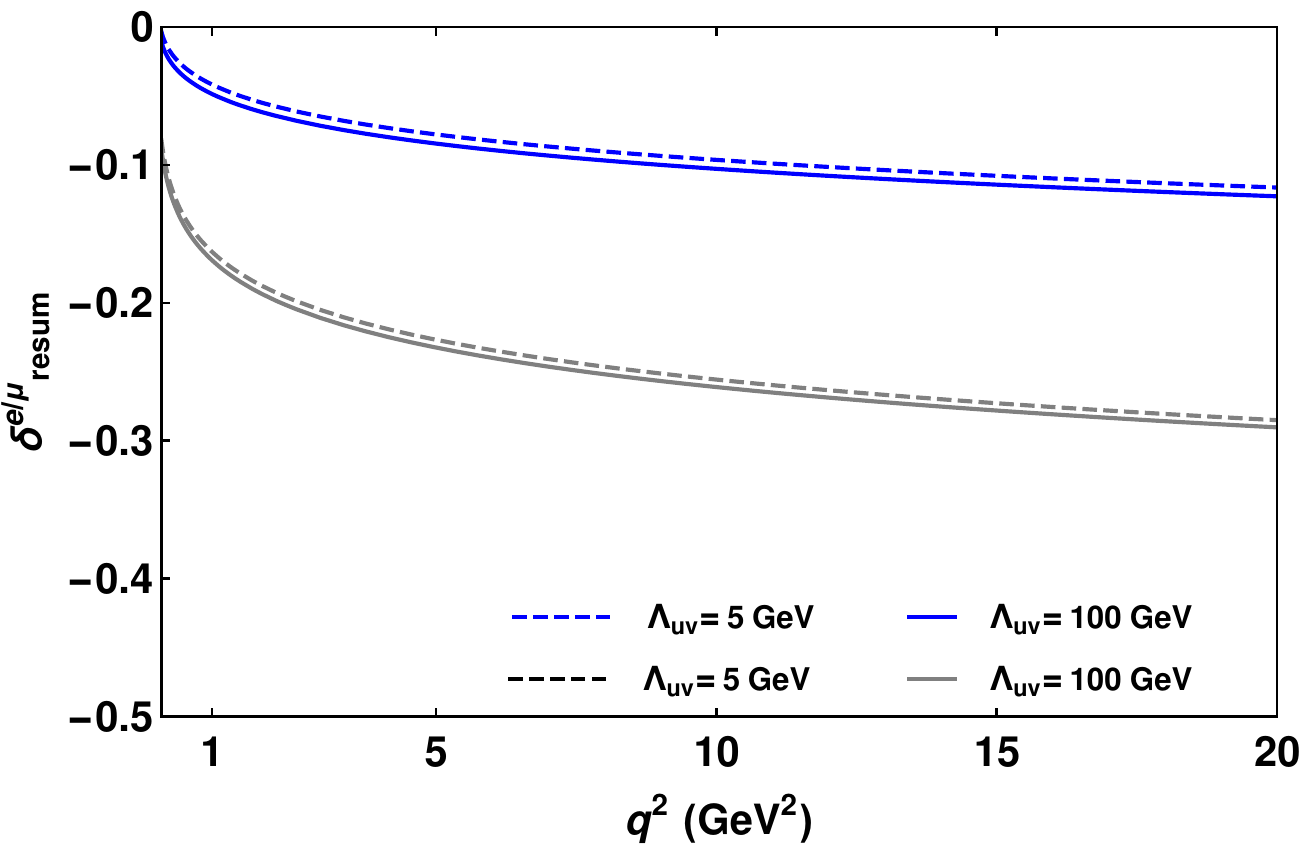}
\caption{The dependence of the QED correction factors $\delta^\ell$ and $\delta^\ell_{\rm
    resum}$ on the UV-cutoff scale $\Lambda_{\rm uv}$ for a fixed $\Delta E_s = 50$ MeV. The solid and dashed grey lines are for electrons, whereas the solid and dashed blue lines are for muons.\label{fig:UV_cutoff}} 
\end{center}
\end{figure}

Finally, we turn to the UV cutoff $\Lambda_{\rm uv}$. This appears through the
Passarino-Veltman functions, and would, ideally, have disappeared were
we able to do an all-order calculation. Nonetheless, the dependence of
the corrections $\delta^\ell \, (\delta^\ell_{\rm resum})$ on this
quantity is very small as is shown in
Fig.\,\ref{fig:UV_cutoff}. Furthermore, the dependence is virtually
independent of the lepton mass, and thus would not affect the
determination of ratios such as $R_{K^{(*)}}$.

\section{Observables}\label{sec:obsv}
With the corrected differential distributions factoring into a
product of the ``tree''-level expression and a $q^2$-dependent
correction (see eqns.\eqref{eq:qedcorr} and \eqref{eq:correxp}), the
actual numerical values for the former were of no consequence until
now. A comparison with the data, however, needs the use of precise
values. To this end, for this analysis, the Wilson coefficients are
  taken from ref. \cite{Detmold:2016pkz} and their expressions and values
  are given in Appendix \ref{app:effWCs}.  The long-distance
  contributions in $H_1\to H_2$ transition matrix elements
  corresponding to different Dirac structures are conventionally
  parametrized in terms of form factors. The HPQCD collaboration
  recently presented a lattice QCD calculation~\cite{Parrott:2022rgu}
  of the $B\to K$ form factor, which is an update of the FNAL/MILC
  calculation presented in ref. \cite{Bailey:2015dka}. We use the
  result from an updated fit \cite{Becirevic:2023aov} that combines
  the HPQCD result \cite{Parrott:2022rgu} with the FNAL/MILC result
  \cite{Bailey:2015dka} in a procedure adopted by the FLAG
  collaboration \cite{FlavourLatticeAveragingGroupFLAG:2021npn}. The
  $B\to K^\ast$ form factors have been
  calculated using light-cone sum
  rules \cite{Bharucha:2015bzk} for low dilepton invariant mass
  squared $q^2$ and using lattice QCD results \cite{Horgan:2013hoa} at high
  $q^2$, and we use a combined fit 
  \cite{Bharucha:2015bzk} to both. The $\Lambda_b\to \Lambda$ and
  $\Lambda_b\to \Lambda^\ast$ form factors as obtained using
  lattice QCD calculations are given in
  refs.~\cite{Detmold:2016pkz} and \cite{Meinel:2020owd, Meinel:2021mdj},
  respectively. All form factors and other parameters relevant to our
  analysis are listed in Appendix~\ref{app:input}.
  
  Before comparing with data, we note that the
    published measurements are partially corrected for
    final-state radiation using
    PHOTOS~\cite{Barberio:1990ms,Golonka:2005pn,Davidson:2010ew}, with
    $q^2$ defined before the emission of
    radiation~\cite{LHCb:2022zom}. PHOTOS reproduces the universal
    soft and collinear logarithms that dominate $\delta^\ell$: for
    $R_K$ it agrees with analytic calculations within one per
    cent~\cite{Bordone:2016gaq}, and for $\bar B^0\to\bar
    K^0\ell^+\ell^-$ at the few per-mille level,
    with resonance-interference effects small for $q^2<6~{\rm
      GeV}^2$~\cite{Isidori:2022bzw}. The corrections presented below
    are therefore relative to a non-radiative computation for a given
    $\Delta E_s$, and are not shifts to be applied to the SM
    predictions when comparing with data; we expect the same to hold
    for the baryonic modes.

\begin{figure}[ht!]
\begin{center}
\includegraphics[scale=0.33]{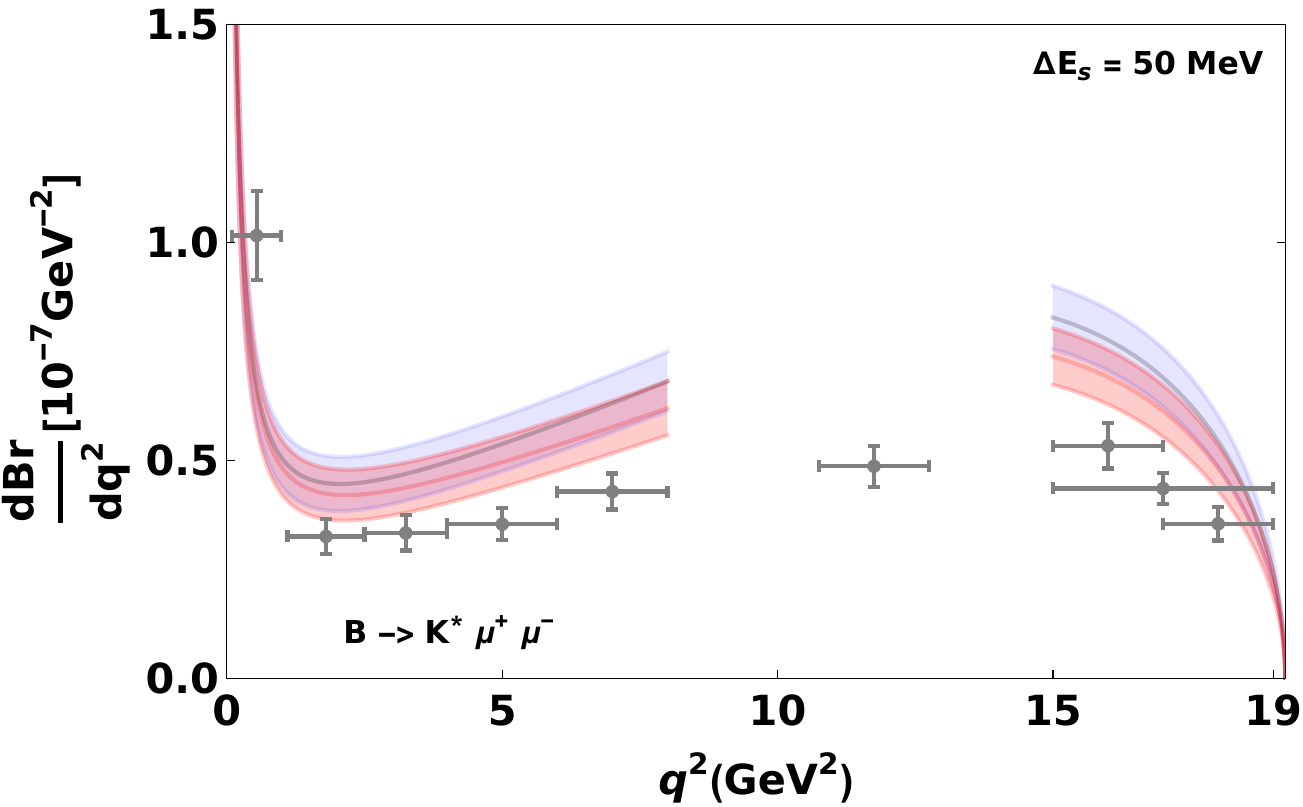}
\includegraphics[scale=0.33]{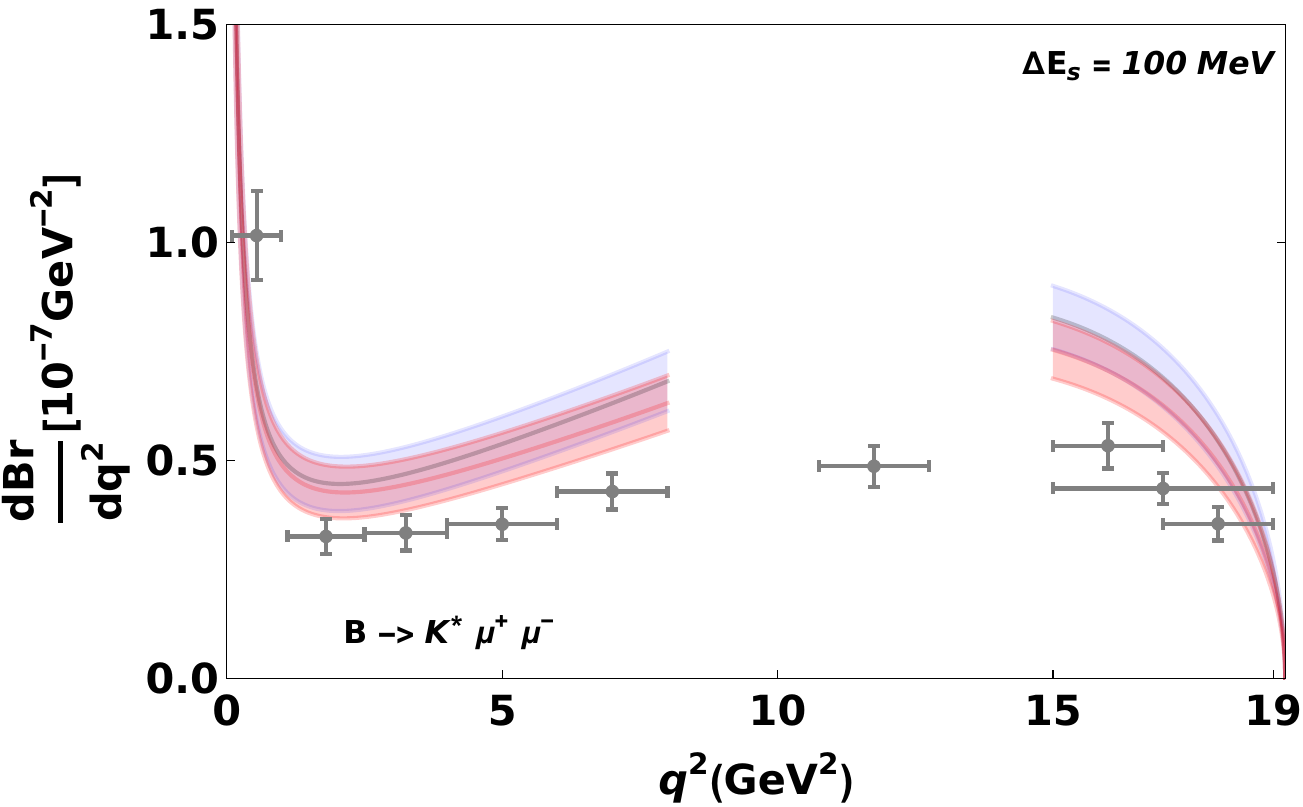}
\includegraphics[scale=0.33]{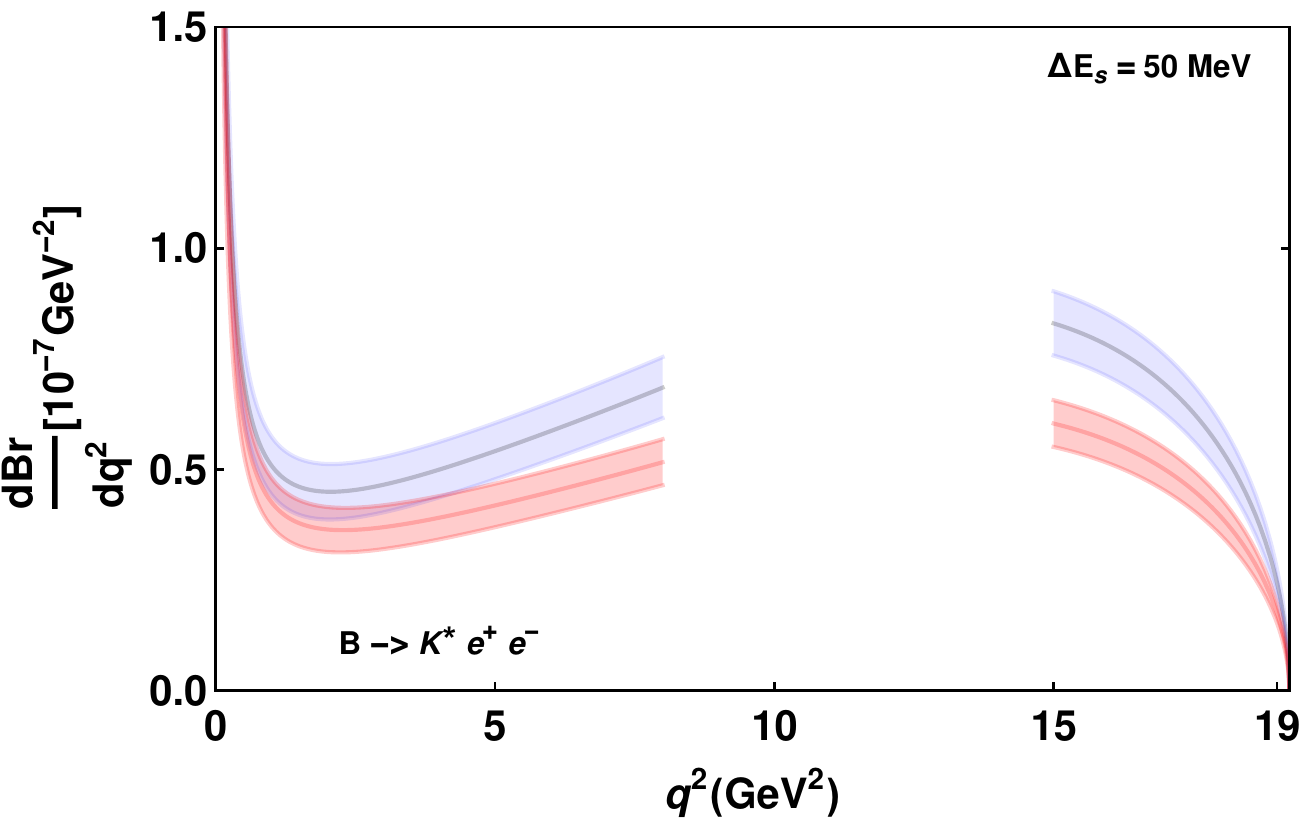}
\includegraphics[scale=0.33]{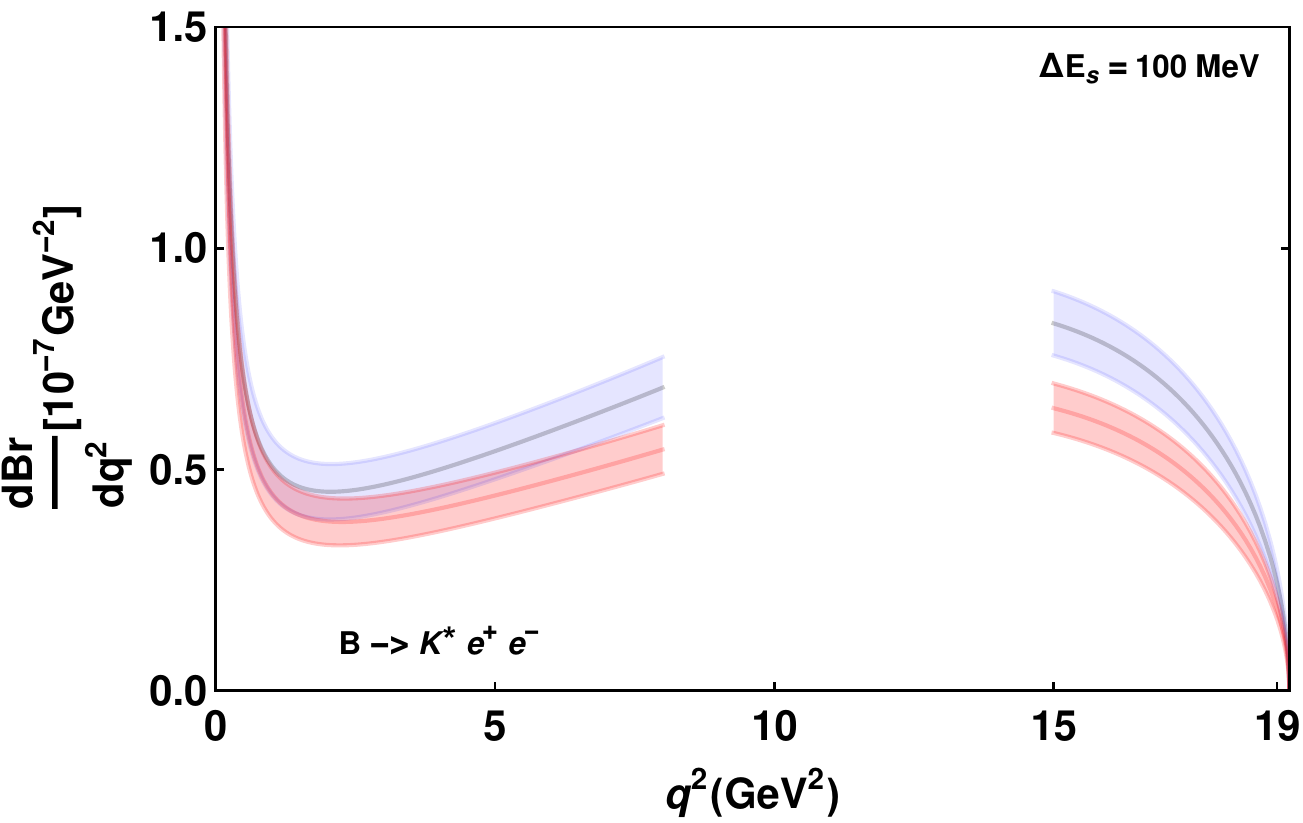}
\caption{Variation of the normalized differential decay width with respect to the
  dilepton invariant mass squared $q^2$ for $B\to K^\ast \mu^+ \mu^-$
  (upper panel) and $B \to K^\ast e^+ e^-$ (lower panel).  The blue
  and red shaded regions correspond to the LO and the NLL predictions
  for $\Delta E_s = \{50,100\}$ MeV, and include both statistical
  and systematic uncertainties.  Also shown (with error bars) are the average differential
 branching ratio data in different $q^2$-bins presented by the LHCb~\cite{LHCb:2016ykl}.  \label{fig:dBrbkst}}
	\end{center}
\end{figure}
\subsection{\boldmath $B\to K^* \ell^+ \ell^-$}
To begin with, we consider $B\to K^* \ell^+ \ell^-$, for both $\ell =
e, \mu$.  In Fig.~\ref{fig:dBrbkst}, we display the differential decay
width (normalized by the total $B$-width $\Gamma_B$) for this mode,
separately for $\ell = e, \mu$. The widths of the bands are obtained
by compounding the uncertainties in the form factors including the
correlations in the same. For the muonic mode, we also display the
experimental data from the LHCb~\cite{LHCb:2016ykl}. Note that we omit
the theoretical curves within the range $8~{\rm GeV}^2 \lsim q^2 \lsim
15~{\rm GeV}^2$. The presence of multiple resonances in that mass
range essentially invalidates any simple formulation; while bound
states can, in principle, be included in the calculations, the
attendant theoretical uncertainties are too large. Note further that
the sharp peak at small $q^2$ is almost at the kinematical edge and is
of limited consequence in the comparison with the binned data.
Finally, it is worth pointing out that the negative correction to the
differential decay width actually brings the theoretical production
{\em closer to the data.} This apparent improvement
  should not be over-interpreted, since the data are partially corrected for final-state radiation (see
  section~\ref{sec:obsv}).

\begin{table}
\begin{scriptsize}
\begin{center}
\setlength{\tabcolsep}{3pt}
\begin{tabular}{| c | c | c | c | c | c | c|}
\hline
&
& \multicolumn{5}{c|}{$B_{\rm bin} \times 10^{7}$}
\\
\hline
& & & \multicolumn{2}{|c|}{$\Delta E_s = 50$ MeV}
& \multicolumn{2}{c|}{$\Delta E_s = 100$ MeV}\\
\cline{4-7}
  $\makecell{\rm Bin \\ (\rm GeV^2)}$
& Decay modes 
& Tree Level & NLO & NLL & NLO & NLL 
\\
\hline
$[0.1,1.1]$ & $B \to K^\ast \mu^+ \mu^- $ &0.7876 $\pm 0.0912$ 
            & 0.7675 $\pm$ 0.0889 & $ 0.7677 \pm 0.0889$
            & 0.7740 $\pm$ 0.0896 & $ 0.7741 \pm 0.0896$  
\\
            & $B \to K^\ast e^+ e^- $ &0.8002 $\pm$ 0.0928   
            & 0.6925 $\pm$ 0.0801     
            & 0.6924 $\pm$ 0.0803 & 0.7276 $\pm$ 0.0846 & 0.7285 $\pm$ 0.0845
\\
\hline
$[1.1,6.0]$ & $B \to K^\ast \mu^+ \mu^- $ & 2.4232$\pm$ 0.3023   
            & 2.2528 $\pm$ 0.2813     & $2.2578 \pm 0.2819$  
            & 2.2905 $\pm$ 0.2860 & $2.2932 \pm 0.2863$
\\
            & $B \to K^\ast e^+ e^- $ &2.4312 $\pm$ 0.3031   
            & 1.8674 $\pm$ 0.2333 & 1.9196 $\pm$ 0.2397
            & 1.9914 $\pm$ 0.2487 & 2.0200 $\pm$ 0.2522 
\\
\hline
$[15,19]$ & $B \to K^\ast \mu^+ \mu^- $ & 2.5820 $\pm$ 0.2502    
            & 2.2802 $\pm$ 0.2209 & $2.2954 \pm 0.2224 $ 
            & 2.3342 $\pm$ 0.2262 & $2.3440 \pm 0.2271 $
\\
            & $B \to K^\ast e^+ e^- $ &2.5863 $\pm$ 0.2506    
            & 1.7607 $\pm$ 0.1705 & 1.8690 $\pm$ 0.1811   
            & 1.9064 $\pm$ 0.1847 & 1.9773 $\pm$ 0.1916 
\\
\hline

\end{tabular}

\end{center}
\end{scriptsize}

\caption{Normalized bin-wise decay widths (see eqn.\eqref{eq:B_bin})
  for the decay $B \to K^\ast \ell^+ \ell^-$.}
\label{tab:Kstar}
\end{table}
While, for the sake of clarity, in Fig.~\ref{fig:dBrbkst}, we have
displayed the results only for the NLL calculated
with one specific choice
for $\Delta E_s$, much the same is true for other choices for $\Delta
E_s$. The NLO results are further suppressed, though. To quantify
this, and to afford an easier comparison with published data, we
introduce the binned decay width through
\begin{equation}
  B_{\rm bin} = \tau_B \int_{\rm Bin} dq^2 \, \frac{d \Gamma}{d q^2} \ ,
\label{eq:B_bin}
\end{equation}
where $\tau_B$ is the lifetime of the meson. In Table \ref{tab:Kstar},
we list $B_{\rm bin}$ for three $q^2$-ranges often used to report
experimental data \cite{LHCb:2016ykl}. The integration over the bins serves to smoothen
out some of the $q^2$ dependence in $\delta^\ell$ and
$\delta^\ell_{\rm resum}$. Nonetheless, the relative sizes of the
integrated widths do show the suppressions with respect to the tree-level results as expected from the analysis
in Section \ref{sec:num} and seen explicitly in Fig.\,\ref{fig:dBrbkst}.
For $B\to K^\ast\mu^+\mu^-$, this reduction is about $1-5\%$ at
low-$q^2$ and between $8-11\%$ at high-$q^2$. For $B\to K^\ast e^+ e^-$,
on the other hand, these corrections are $8-15\%$ at low-$q^2$ and
roughly $23-28\%$ at high-$q^2$. 

What is perhaps even more interesting is the value of the ratio
$R_{K^*}$ as listed in Table \ref{tab:rkstar}. Note that the
  uncertainty in the ratio is much smaller than those in the
  individual partial decay widths (binned or unbinned). This is but a
  reflection of the fact that with $m_\ell^2 \ll m_B^2 - m_{K^*}^2$,
  the arguments of the form factors in the two decay processes are
  virtually the same and, hence, the uncertainties in the two decays
  are entirely correlated. With the electronic mode suffering a
larger negative correction as compared to the muonic mode, the
corrected $R_{K^*}$ increases from the tree-level value.
This increase, however, is relative to a non-radiative
computation and is not to be compared directly with
eqns.(\ref{data:lowqsq}) and (\ref{data:centqsq}) (see
section~\ref{sec:obsv}).

\begin{table}[H]
\begin{footnotesize}
\begin{center}
\setlength{\tabcolsep}{3pt}
\begin{tabular}{| c | c | c | c | c | c|c|c| }
\hline \hline
$\makecell{\rm Bin \\ (\rm GeV^2)}$    &  \makecell{ \rm Tree-level  } &\makecell{$\rm NLO $\\($\Delta E_s = 50$ MeV)} & \makecell{$\rm NLL $\\($\Delta E_s = 50$ MeV)} & \makecell{$\rm NLO$\\($\Delta E_s = 100$ MeV)} & \makecell{$\rm NLL $\\($\Delta E_s = 100$ MeV)} \\
 \hline
 $[0.1,1.1]$ & 0.98429 $\pm 0.00089 $   & 1.10842 $\pm 0.00204 $    & 1.10109 $\pm 0.00184$  & 1.06392 $\pm 0.00188 $ & 1.06265 $\pm 0.00178 $    \\
 
 \hline
 $[1.1,6.0]$ & 0.99667 $\pm 0.00048 $   & 1.20638 $\pm 0.00040 $    & 1.17617 $\pm 0.00043 $  &1.15022 $\pm 0.00039$ & 1.13526 $\pm 0.00042$    \\

 \hline
 $[15,19]$ & 0.99833 $\pm 0.00003 $   & 1.29506 $\pm 0.00009 $    & 1.22814 $\pm 0.00006 $  &1.22444 $\pm 0.00008$ & 1.18543 $\pm 0.00006$      \\

 \hline
 
\end{tabular}
\end{center}
\end{footnotesize}

\caption{The LFU variable $R_{K^\ast}$ for the low-, the central- and
  the high-$q^2$ regions.}
\label{tab:rkstar}
\end{table}  
%

\subsection{\boldmath $B\to K \ell^+ \ell^-$}

\begin{figure}[ht!]
\begin{center}
  \includegraphics[scale=0.33]{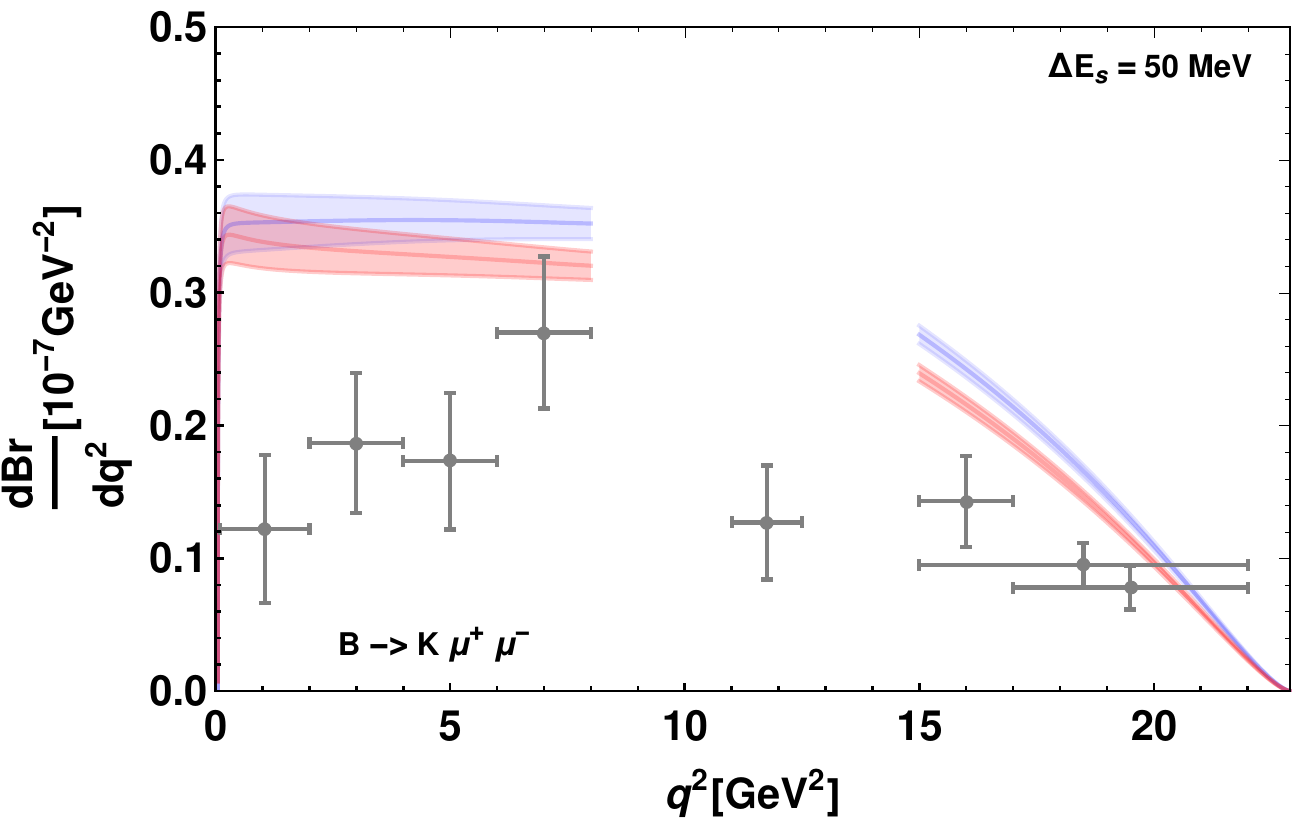}
  \includegraphics[scale=0.33]{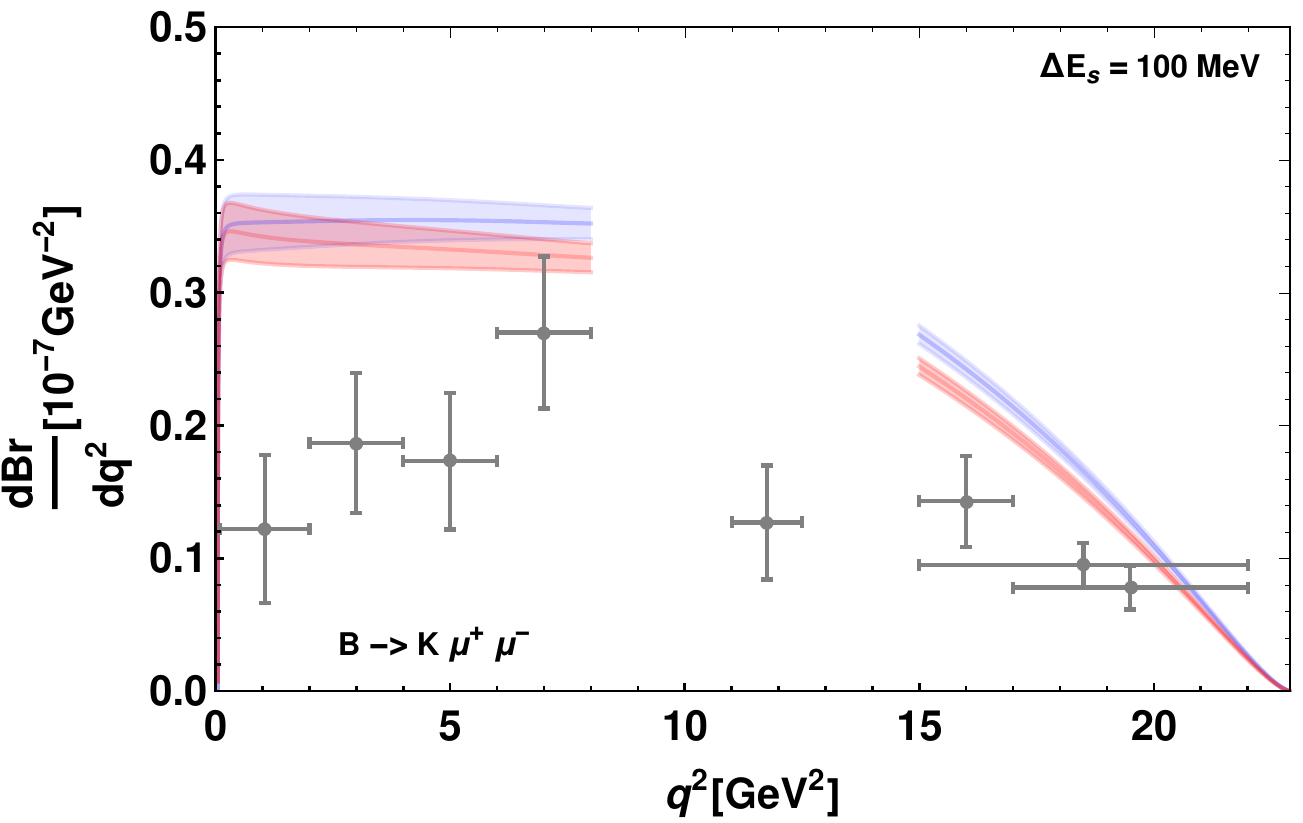}
  \includegraphics[scale=0.33]{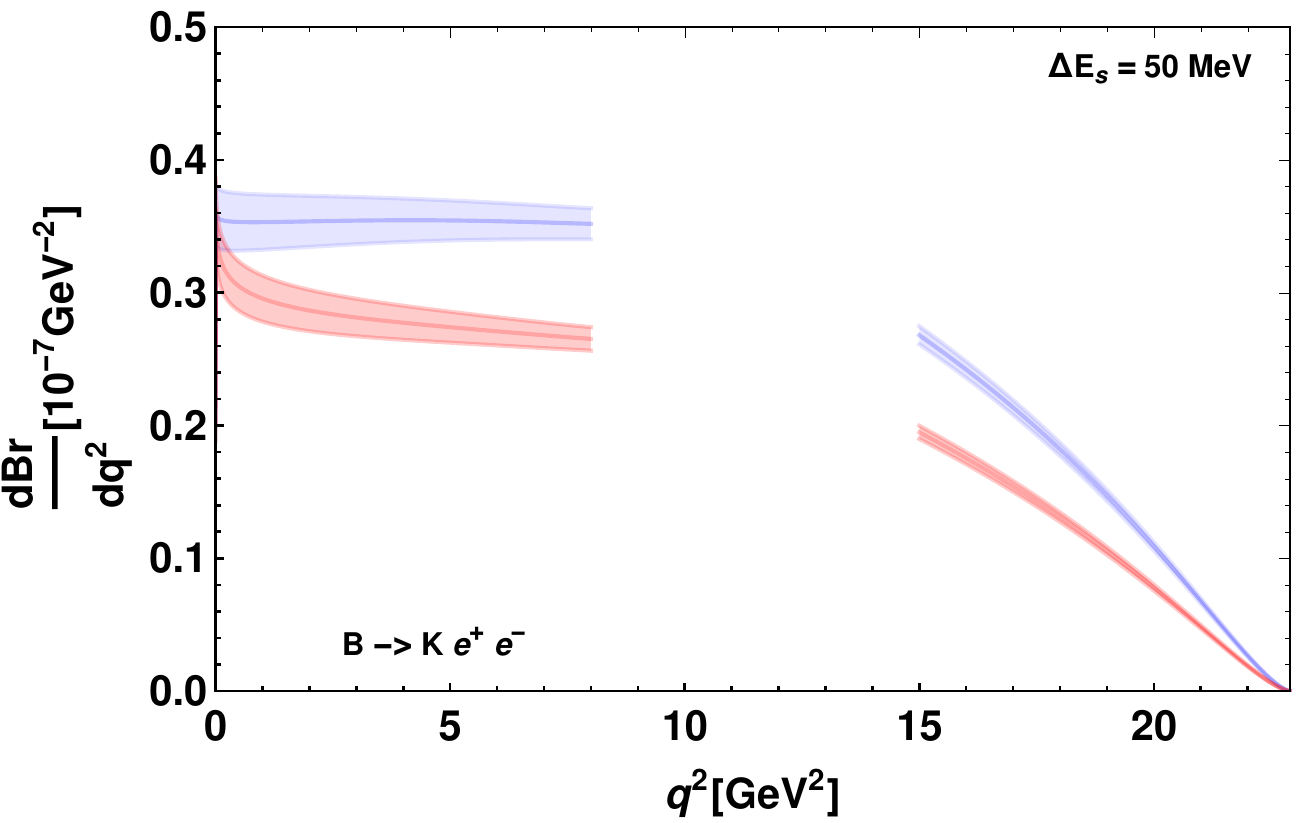}
  \includegraphics[scale=0.33]{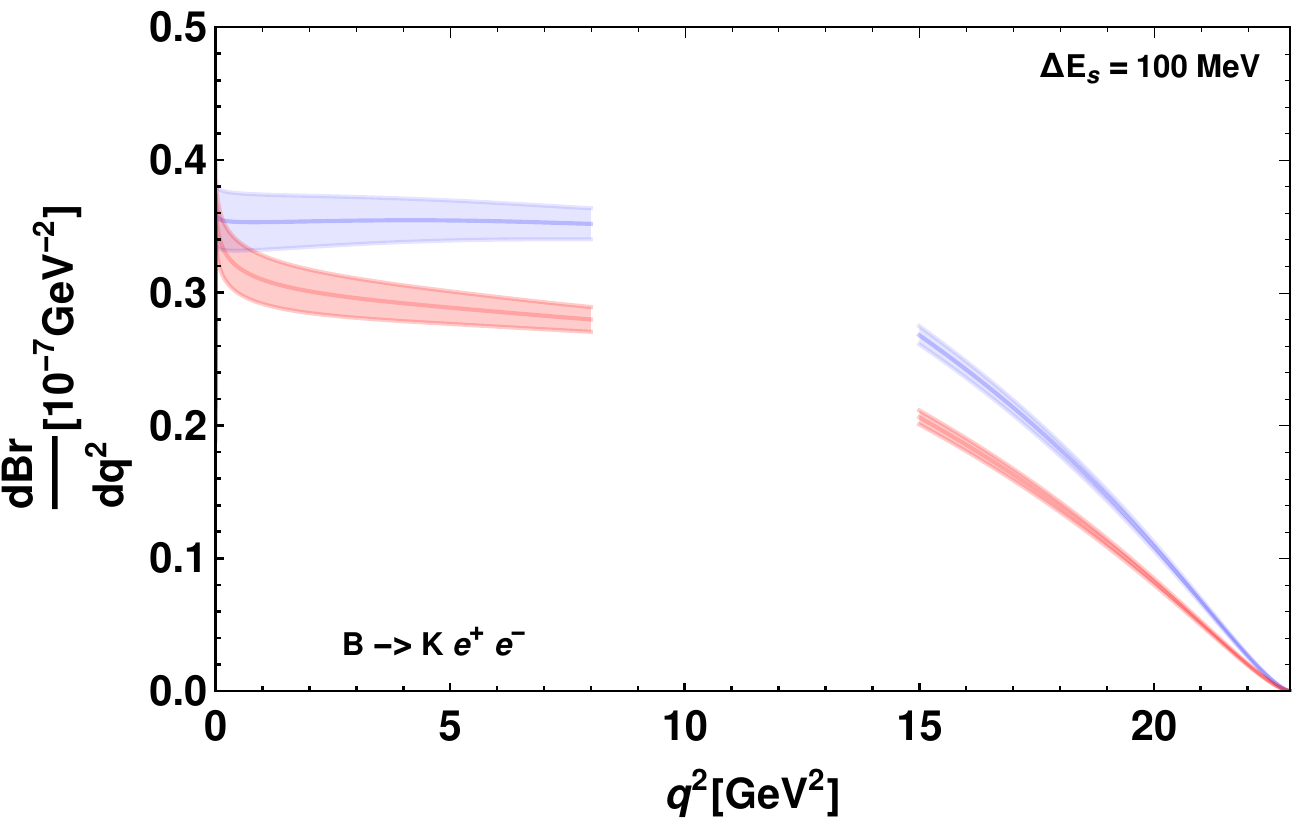}
\end{center}
\caption{As in Fig.\,\ref{fig:dBrbkst}, but for $B \to K \ell^+ \ell^-$
  instead. The experimental data is taken from ref.~\cite{LHCb:2014cxe}.
}
\label{fig:dBrbk}
\end{figure}

From $B\to K^\ast \ell^+ \ell^-$, we now turn to the
sister decay, {\em viz.} $B\to K \ell^+ \ell^-$. In
  Fig.\,\ref{fig:dBrbk}, we display the corresponding differential decay
  distributions. Once again, the theoretical predictions for $8~{\rm
    GeV}^2 \lsim q^2 \lsim 15~{\rm GeV}^2$ are beset with large
  uncertainties (on account of the presence of multiple resonances)
  and are omitted.  At high-$q^2$, the behavior is quite similar to
  that in the case for the decay into $K^\ast$ and can be understood
  from kinematic considerations.  Note, though, the remarkably
  different behavior at small $q^2$ values.  Unlike in the previous
  case, there is no precipitous rise at ultrasmall $q^2$ values. The
  flattened behavior instead, and more importantly, the large
  discrepancy with measurements is well documented and has been
  discussed extensively in the literature. 

What is of immediate concern is the fact that, once again, the NLL (as
also the NLO) curves sit below the leading-order curves (as was
to be expected from the discussions in Section \ref{sec:num}) and
serve to reduce the systematic `deficiency' that
the experimental numbers suffer from (when compared to the LO
predictions). These conclusions are also reflected in the bin-wise
integrated partial widths as listed in Table~\ref{tab:K}.

\begin{table}[H]
\begin{scriptsize}
\begin{center}
\setlength{\tabcolsep}{3pt}
\begin{tabular}{| c | c | c | c | c | c | c|}
\hline
&
& \multicolumn{5}{c|}{$B_{\rm bin} \times 10^{7}$}
\\
\hline
& & & \multicolumn{2}{|c|}{$\Delta E_s = 50$ MeV}
& \multicolumn{2}{c|}{$\Delta E_s = 100$ MeV}\\
\cline{4-7}
  $\makecell{\rm Bin \\ (\rm GeV^2)}$
& Decay modes 
& Tree Level & NLO & NLL & NLO & NLL 
\\
\hline
$[0.1,1.1]$ & $B \to K \mu^+ \mu^- $ &0.3498 $\pm$ 0.0213
            & 0.3393 $\pm$ 0.0207 & $ 0.3394 \pm 0.0207$
            & 0.3425 $\pm$ 0.0209 & $ 0.3425 \pm 0.0209$  
\\
            & $B \to K e^+ e^- $ &0.3523 $\pm$ 0.0215   
            & 0.3009 $\pm$ 0.0183 & 0.3035 $\pm$ 0.0185    
            & 0.3167 $\pm$ 0.0193 & 0.3174 $\pm$ 0.0193
\\
\hline
$[1.1,6.0]$ & $B \to K \mu^+ \mu^- $ & 1.7326 $\pm$ 0.0830   
            & 1.6122 $\pm$ 0.0774     & $ 1.6157 \pm 0.0775$  
            & 1.6390 $\pm$ 0.0786 & $ 1.6408 \pm 0.0787$
\\
            & $B \to K e^+ e^- $ & 1.7321 $\pm$ 0.0829   
            & 1.3332 $\pm$ 0.0640 & 1.3698 $\pm$ 0.0657
            & 1.4213 $\pm$ 0.0682 & 1.4413 $\pm$ 0.0692 
\\
\hline
$[15,19]$ & $B \to K \mu^+ \mu^- $ & 0.8480 $\pm$ 0.0227    
            & 0.7488 $\pm$ 0.0200 & $ 0.7538 \pm 0.0202 $ 
            & 0.7665 $\pm$ 0.0205 & $ 0.7697 \pm 0.0206 $
\\
            & $B \to K e^+ e^- $ & 0.8465 $\pm$ 0.0227    
            & 0.5761 $\pm$ 0.0154 & 0.6116 $\pm$ 0.0164   
            & 0.6238 $\pm$ 0.0167 & 0.6470 $\pm$ 0.0173 
\\
\hline

\end{tabular}

\end{center}
\end{scriptsize}

\caption{Normalized bin-wise decay widths (see eqn.\eqref{eq:B_bin})
  for the decay $B \to K \ell^+ \ell^-$.}
\label{tab:K}
\end{table}

As for the LFU variable $R_K$, once again the uncertainties are
  tiny (see Table \ref{tab:rk}), and for the same reasons as operative
  for $R_{K^*}$. More importantly, the theoretical expectation at NLL
  are significantly larger (than those at the tree-level). As for $R_{K^*}$, this enhancement is relative to a
non-radiative computation and does not imply a tension with the
measured values. Our results are in agreement with \cite{Isidori:2020acz, Mishra:2020orb}.

\begin{table}[H]
\begin{footnotesize}
\begin{center}
\setlength{\tabcolsep}{3pt}
\begin{tabular}{| c | c | c | c | c | c|c|c| }
\hline \hline
$\makecell{\rm Bin \\ (\rm GeV^2)}$    &  \makecell{ \rm Tree-level  } &\makecell{$\rm NLO $\\($\Delta E_s = 50$ MeV)} & \makecell{$\rm NLL $\\($\Delta E_s = 50$ MeV)} & \makecell{$\rm NLO$\\($\Delta E_s = 100$ MeV)} & \makecell{$\rm NLL $\\($\Delta E_s = 100$ MeV)} \\
 \hline
 $[0.1,1.1]$ & 0.99300 $\pm 0.00006 $   & 1.12748 $\pm 0.00004 $    & 1.11833 $\pm 0.00005$  & 1.08153 $\pm 0.00004 $ & 1.07932 $\pm 0.00005 $    \\
 
 \hline
 $[1.1,6.0]$ & 1.00029 $\pm 0.00007 $   & 1.20929 $\pm 0.00007 $    & 1.17949 $\pm 0.00004 $  &1.15315 $\pm 0.00006$ & 1.13847 $\pm 0.00004$    \\

 \hline
 $[15,19]$ & 1.00175 $\pm 0.00007 $   & 1.29977 $\pm 0.00008 $    & 1.23249 $\pm 0.00008 $  &1.22886 $\pm 0.00008$ & 1.18962 $\pm 0.00008$      \\

 \hline
 
\end{tabular}
\end{center}
\end{footnotesize}
\caption{The LFU variable $R_{K}$ for the low-, the central- and
  the high-$q^2$ regions.}
\label{tab:rk}
\end{table}  
%

\subsection{\boldmath $\Lambda_b \to \Lambda \ell^+ \ell^-$ and $\Lambda_b \to \Lambda^* \ell^+ \ell^-$ }
\begin{figure}[ht!]
  \begin{center}
    \includegraphics[scale=0.33]{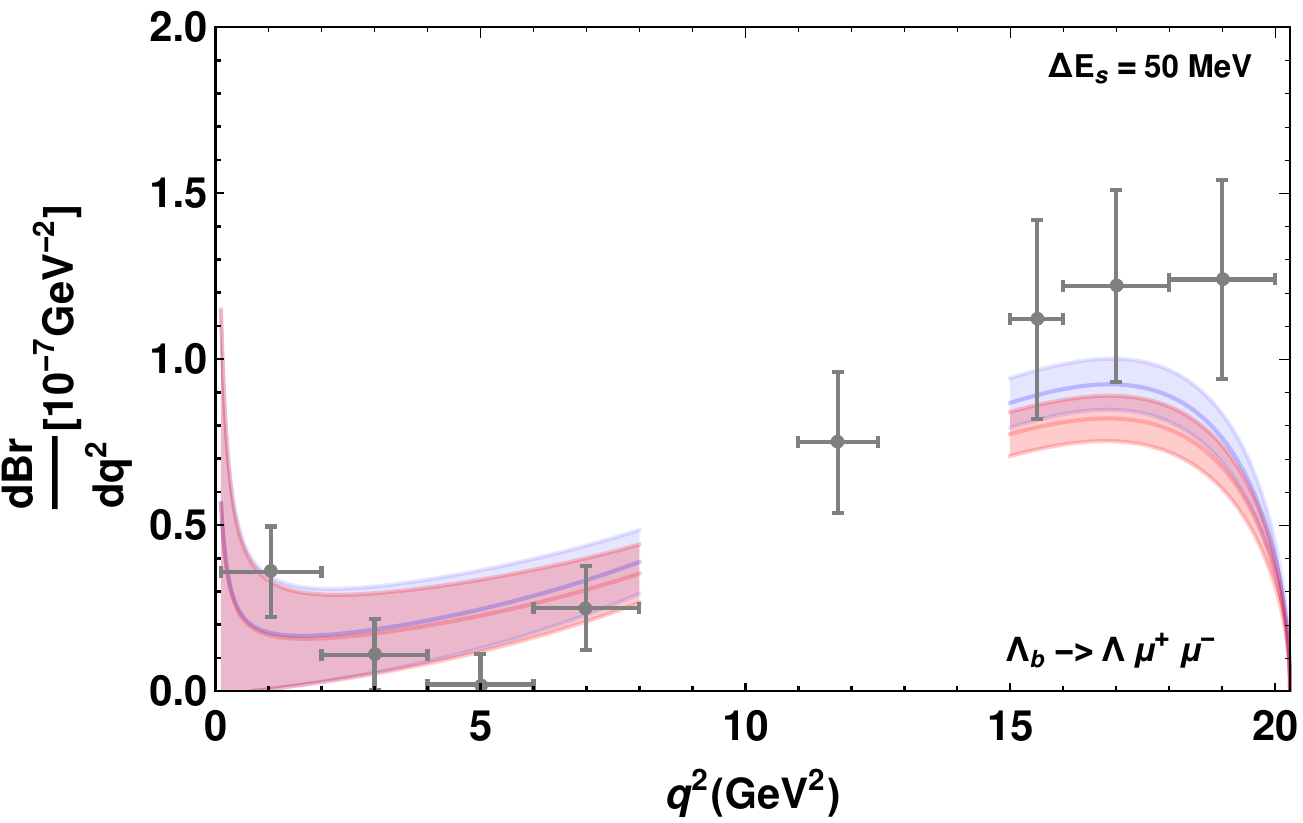}
    \includegraphics[scale=0.33]{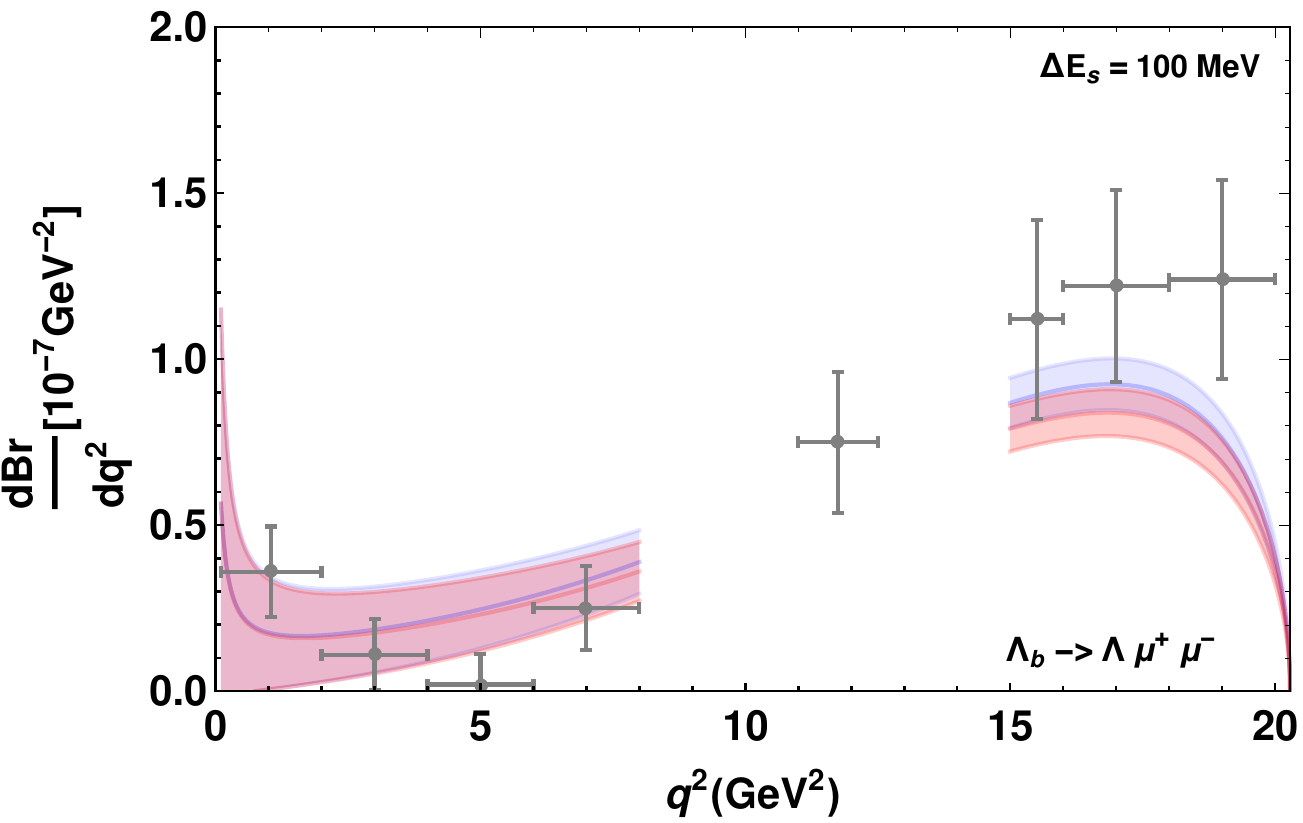}
    \includegraphics[scale=0.33]{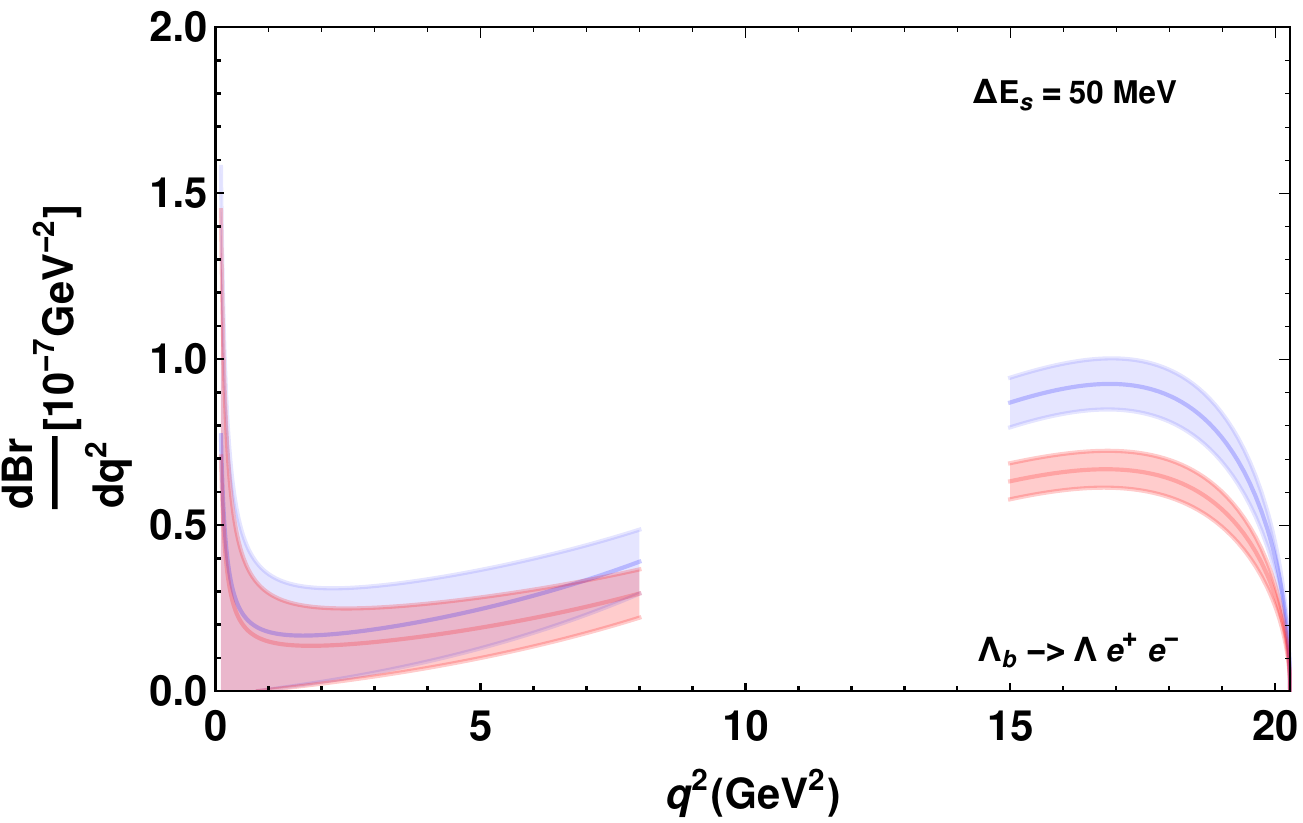}
    \includegraphics[scale=0.33]{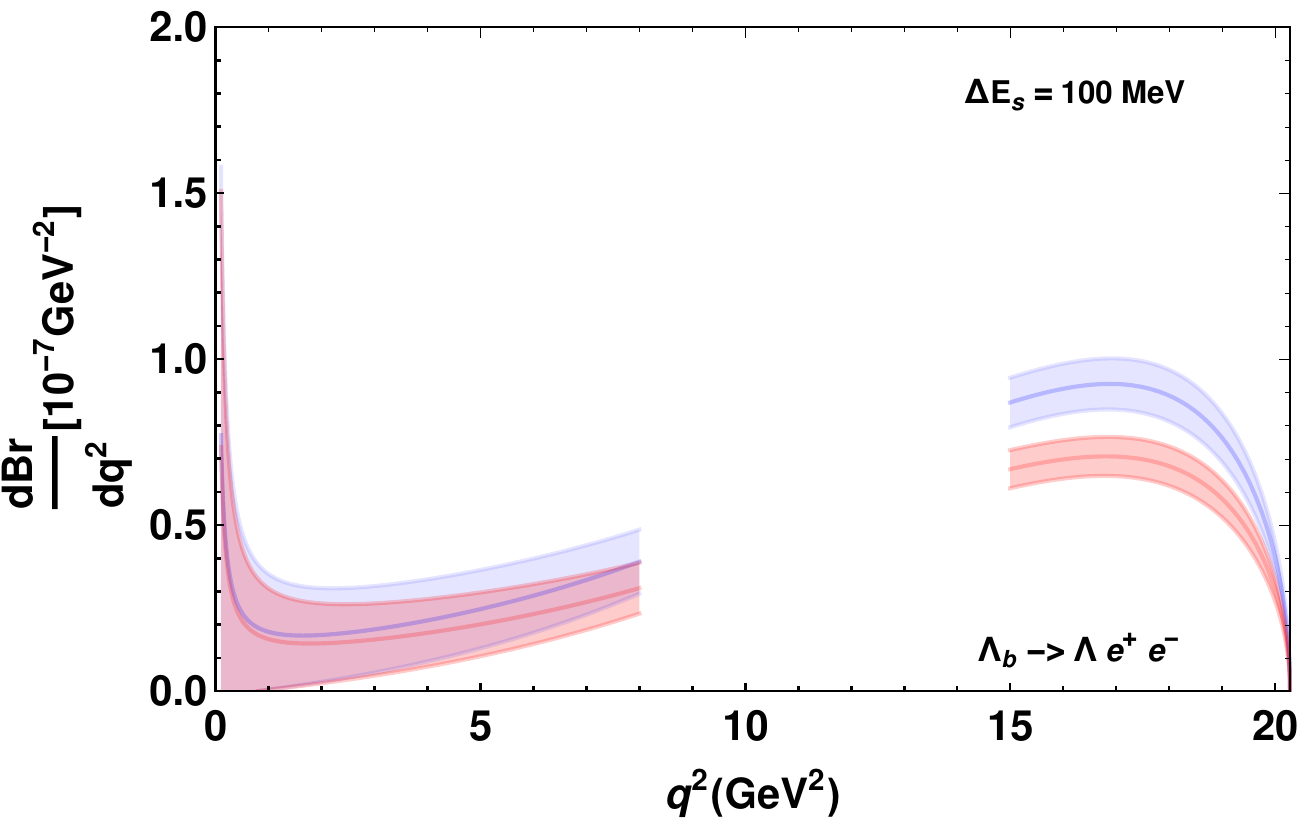}		
  \end{center}
  \caption{As in Fig.\,\ref{fig:dBrbkst}, but for $\Lambda_b \to \Lambda \ell^+ \ell^-$
    instead. The experimental data is taken from ref.\cite{LHCb:2015tgy}.}
  \label{fig:dBrlamda}
\end{figure}

Finally, we move to decays of the baryon $\Lambda_b$ as opposed
  to the mesons that we had dealt with so far. The $\Lambda_b\to \Lambda\mu^+\mu^-$ was first observed by the CDF collaboration \cite{CDF:2011buy}. Subsequently, the LHCb Collaboration studied the decay rate against the dimuon invariant mass squared in \cite{LHCb:2013uqx}, and the angular distribution in \cite{LHCb:2015tgy, LHCb:2018jna}. Theoretical studies in the SM and NP has been performed in \cite{Gutsche:2013pp, Boer:2014kda, Das:2018sms, Blake:2017une,Roy:2017dum,Das:2018iap,Das:2022xjg,Das:2020qws,Meinel:2016grj,Bhattacharya:2019der,Blake:2019guk,Das:2023kch}. Relatively recently the $\Lambda_b\to \Lambda^\ast\ell^+\ell^-$, where $\Lambda^\ast$ is a spin-3/2 hadron, has been extensively studied \cite{Descotes-Genon:2019dbw,Das:2020cpv,Amhis:2020phx,Meinel:2020owd,Meinel:2021mdj,Amhis:2022vcd,Li:2022nim,Biswas:2022fyb,Das:2023kch} in the SM and for various new physics scenarios. The LHCb has also reported their first measurements of the $\Lambda_b\to \Lambda^\ast\ell^+\ell^-$ branching fraction in \cite{LHCb:2023ptw}. 

While the structure of the QED corrections in $\Lambda_b\to \Lambda^{(\ast)}\ell^+\ell^-$ remains rather
  similar, the basic (``tree''-level) decay matrix element changes
  substantially and so do the form factors. In Fig.\,\ref{fig:dBrlamda}, 
we display the differential branching ratio $d\mathcal{B}/dq^2$ for the $\Lambda_b \to \Lambda\ell^+\ell^-$ decay as a function of $q^2$. As earlier, we choose to compare the tree-level (blue) and
  the NLL (red) expectations, and compute the latter for two different
  choices of $\Delta E_s$. 

Several comments need to be made at this stage:
  \begin{itemize}
    \item The relative errors in the data are understandably larger (than those for $B \to K^{(*)}\ell^+\ell^-$ decays) owing to the smaller $\Lambda_b$ production cross sections. \item More strikingly, the errors in the theoretical
    expectations are much larger too.   These are largely dominated by the
    uncertainties in the form factors.
  \item In view of the large uncertainties, we have used two sets of
    form factor parametrizations. The ``nominal'' set of parameters
    \cite{Detmold:2016pkz} (and their correlation matrix) are used to
    compute the central values and statistical uncertainties. In
    addition, the ``higher-order'' parameters and their correlation
    matrix are combined with the nominal parameters to compute the
    systematic uncertainties in a fully correlated manner. This
    inclusion of the systematic uncertainties (in addition to the
    statistical uncertainties) in the form factors leads to the
    significantly larger errors in the differential decay width
    distribution.
  \item Not surprisingly, the relative errors (theoretical) are much smaller
    for large-$q^2$ values owing to the more accurate lattice-QCD results
    that are applicable there.
  \item As far as the lowest order predictions go, at low-$q^2$, they
    agree well with the experimental data \cite{LHCb:2015tgy}, whereas
    at high-$q^2$, the data exceeds the predictions.
  \item With $\delta^\ell_{\rm resum}$ being negative, the QED
    corrected differential branching ratios are lower than the
    uncorrected values. While they are still consistent with the data
    at small $q^2$, for large $q^2$ the disagreement with the data
    worsens. As the LHCb
    measurement~\cite{LHCb:2015tgy} likewise corrects for final-state
    radiation, this should not be over-interpreted.
\item Once again, the corrections are negative and larger for
  electrons as compared to muons. The corrections increase with $q^2$
  for both cases. Similarly, it decreases with increasing $\Delta E_s$
  (for a fixed value of $q^2$)
\item In particular, 
for $\Lambda_b \to \Lambda \mu^+\mu^-$ (see Fig.\,\ref{fig:dBrlamda}), the NLL partial widths differ
from the LO ones by $1\text{-} 5\%$ at low-$q^2$ and $5\text{-}
  10\%$ at high-$q^2$ when $\Delta E_s = 50$ MeV is used. For
  $\Delta E_s = 100$ MeV, the corrections are marginally smaller.
  For $\Lambda_b\to \Lambda e^+ e^-$, on the other hand, the corresponding differences 
  are $8\text{-} 18\%$ ($5\text{-} 15\%$) at low-$q^2$, and 
  $22\text{-} 27\%$ ( $18\text{-} 23\%$) at high-$q^2$ when $\Delta E_s = 50 (100)$
  MeV is used. 
  \end{itemize}

\begin{table}
%
\begin{footnotesize}
\begin{center}
\setlength{\tabcolsep}{3pt}
\begin{tabular}{| c | c | c | c | c | c | c|}
\hline
&
& \multicolumn{5}{c|}{$B_{\rm bin} \times 10^{7}$}
\\
\hline
& & & \multicolumn{2}{|c|}{$\Delta E_s = 50$ MeV}
& \multicolumn{2}{c|}{$\Delta E_s = 100$ MeV}\\
\cline{4-7}
  $\makecell{\rm Bin \\ (\rm GeV^2)}$
& Decay modes 
& Tree Level & NLO & NLL & NLO & NLL 
\\
\hline
$[0.1,1.1]$ & $\Lambda_b \to \Lambda \mu^+ \mu^- $ &0.247 $\pm 0.249$ 
            & 0.241 $\pm$ 0.242 & $ 0.241 \pm 0.243$
            & 0.243 $\pm$ 0.245 & $ 0.243 \pm 0.245$  
\\
            & $\Lambda_b \to \Lambda e^+ e^- $ &0.265 $\pm$ 0.270   
            & 0.229 $\pm$ 0.234 & 0.231 $\pm$ 0.235
            & 0.241 $\pm$ 0.246 & 0.241 $\pm$ 0.246  

\\
\hline
$[1.1,6.0]$ & $\Lambda_b \to \Lambda \mu^+ \mu^- $ & 1.017 $\pm$ 0.630   
            & 0.946 $\pm$ 0.583     & $ 0.949 \pm 0.584$  
            & 0.960 $\pm$ 0.597 & $ 0.962 \pm 0.597$
\\
            & $\Lambda_b \to \Lambda e^+ e^- $ & 1.020 $\pm$ 0.634   
            & 0.783 $\pm$ 0.486 & 0.805 $\pm$ 0.499
            & 0.833 $\pm$ 0.522 & 0.846 $\pm$ 0.529 
\\
\hline
$[15,19]$ & $\Lambda_b \to \Lambda \mu^+ \mu^- $ & 3.561 $\pm$ 0.302    
            & 3.143 $\pm$ 0.266 & $ 3.164 \pm 0.268 $ 
            & 3.217 $\pm$ 0.273 & $ 3.231 \pm 0.274 $
\\
            & $\Lambda_b \to \Lambda e^+ e^- $ & 3.565 $\pm$ 0.302    
            & 2.424 $\pm$ 0.205 & 2.574 $\pm$ 0.218   
            & 2.625 $\pm$ 0.222 & 2.723 $\pm$ 0.231 
\\
\hline

\end{tabular}

\end{center}

\end{footnotesize}

\caption{Normalized bin-wise decay widths (see eqn.\eqref{eq:B_bin})
  for the decay $\Lambda_b \to \Lambda \ell^+ \ell^-$.}
\label{tab:Lambda}
\end{table}
  Given the relative paucity of data and the consequent
  fluctuations, it would be useful to consider binned data as in the
  case of $B$-decays. Unfortunately, though, no such integrated data
  is published and, hence, there are no natural guidelines for
  binning. Given this situation, we adopt the same binning as for the
  $B$-decays, and, in Table \ref{tab:Lambda}, we list the
  corresponding bin-wise numbers. The large error bars for the low--
  and central--$q^2$ bins are reflective of the first two points
  discussed above. Notwithstanding the errors, it is still worthwhile
  to calculate the ratio of the muonic and electronic decays, for the
  errors in the individual channels are highly correlated and would largely
  cancel in the ratio.

\begin{table}
\begin{footnotesize}
\begin{center}
\setlength{\tabcolsep}{3pt}
\begin{tabular}{| c | c | c | c | c | c|c|c| }
\hline \hline
$\makecell{\rm Bin \\ (\rm GeV^2)}$    &  \makecell{ \rm Tree-level  } &\makecell{$\rm NLO $\\($\Delta E_s = 50$ MeV)} & \makecell{$\rm NLL $\\($\Delta E_s = 50$ MeV)} & \makecell{$\rm NLO$\\($\Delta E_s = 100$ MeV)} & \makecell{$\rm NLL $\\($\Delta E_s = 100$ MeV)} \\
 \hline
 $[0.1,1.1]$ & 0.93388 $\pm 0.00923 $   & 1.05177 $\pm 0.01098 $    & 1.04469 $\pm 0.01080$  & 1.00962 $\pm 0.01049 $ & 1.00834 $\pm 0.01042 $    \\
 
 \hline
 $[1.1,6.0]$ & 0.99719 $\pm 0.00175 $   & 1.20897 $\pm 0.00423 $    & 1.17805 $\pm 0.00343 $  &1.15249 $\pm 0.00380$ & 1.13709 $\pm 0.00330$    \\

 \hline
 $[15,19]$ & 0.99881 $\pm 0.00003 $   & 1.29656 $\pm 0.00010 $    & 1.22918 $\pm 0.00006 $  &1.22575 $\pm 0.00009$ & 1.18643 $\pm 0.00006$      \\

 \hline
 
\end{tabular}
\end{center}
\end{footnotesize}

\caption{The LFU variable $R_{\Lambda}$ for the low-, the central- and
  the high-$q^2$ regions..}
\label{tab:rLambda}
\end{table}
Defining a LFU variable $R_\Lambda$ (in parallel with $R_{K}$
  and $R_{K^*}$), in Table \ref{tab:rLambda} we list its value for the
  aforementioned three bins. Once again, the NLL value for $R_\Lambda$
  is significantly larger than that at the tree-level (with the NLO
  value being even larger). This could be of great interest once the data
  is robust enough for such measurements to be meaningful. 

\begin{figure}[ht!]
  \begin{center}
    \includegraphics[scale=0.32]{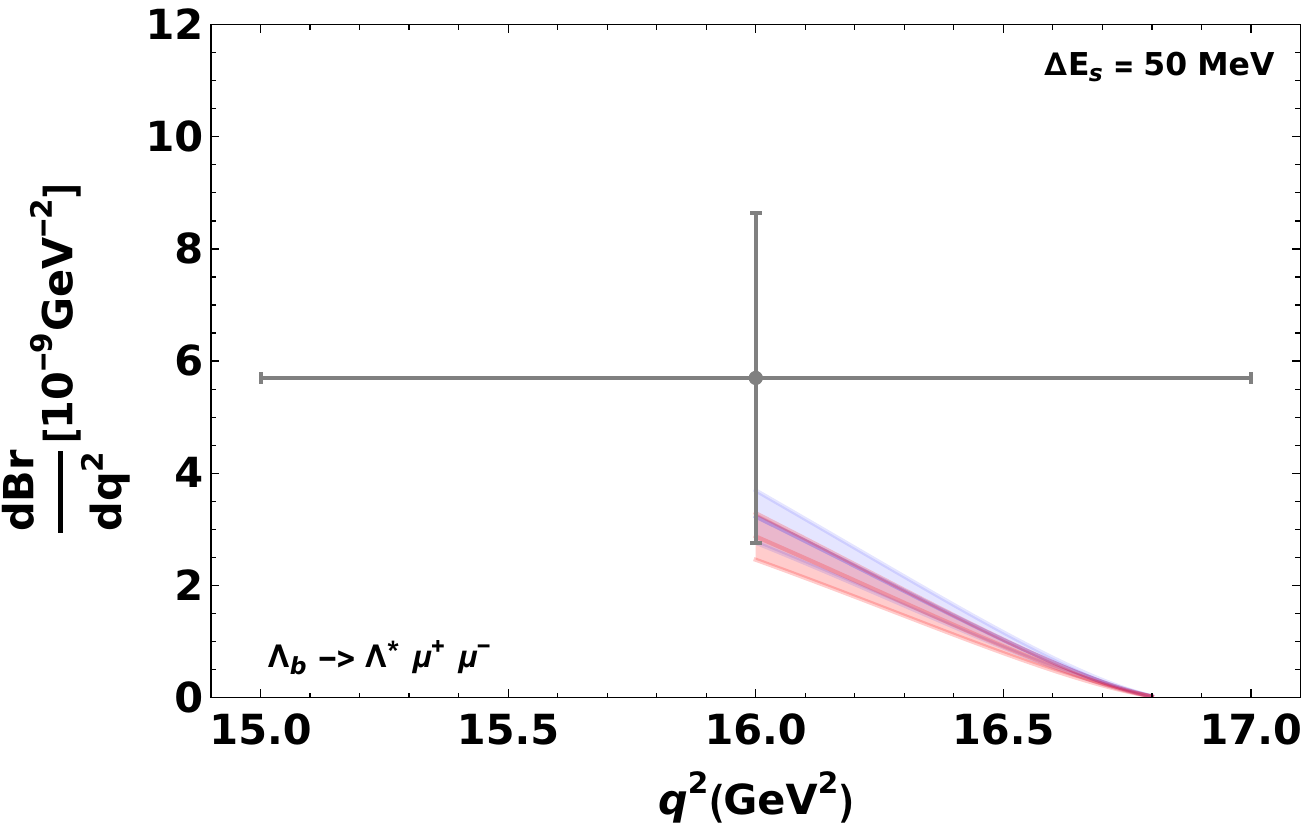}
    \includegraphics[scale=0.56]{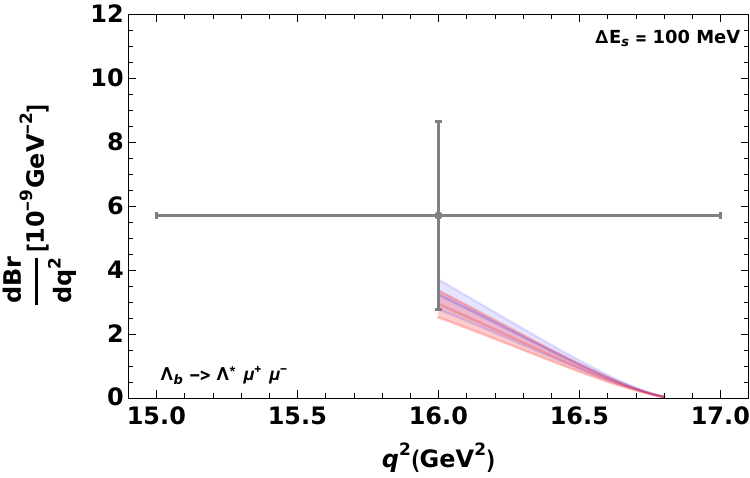}
    \includegraphics[scale=0.33]{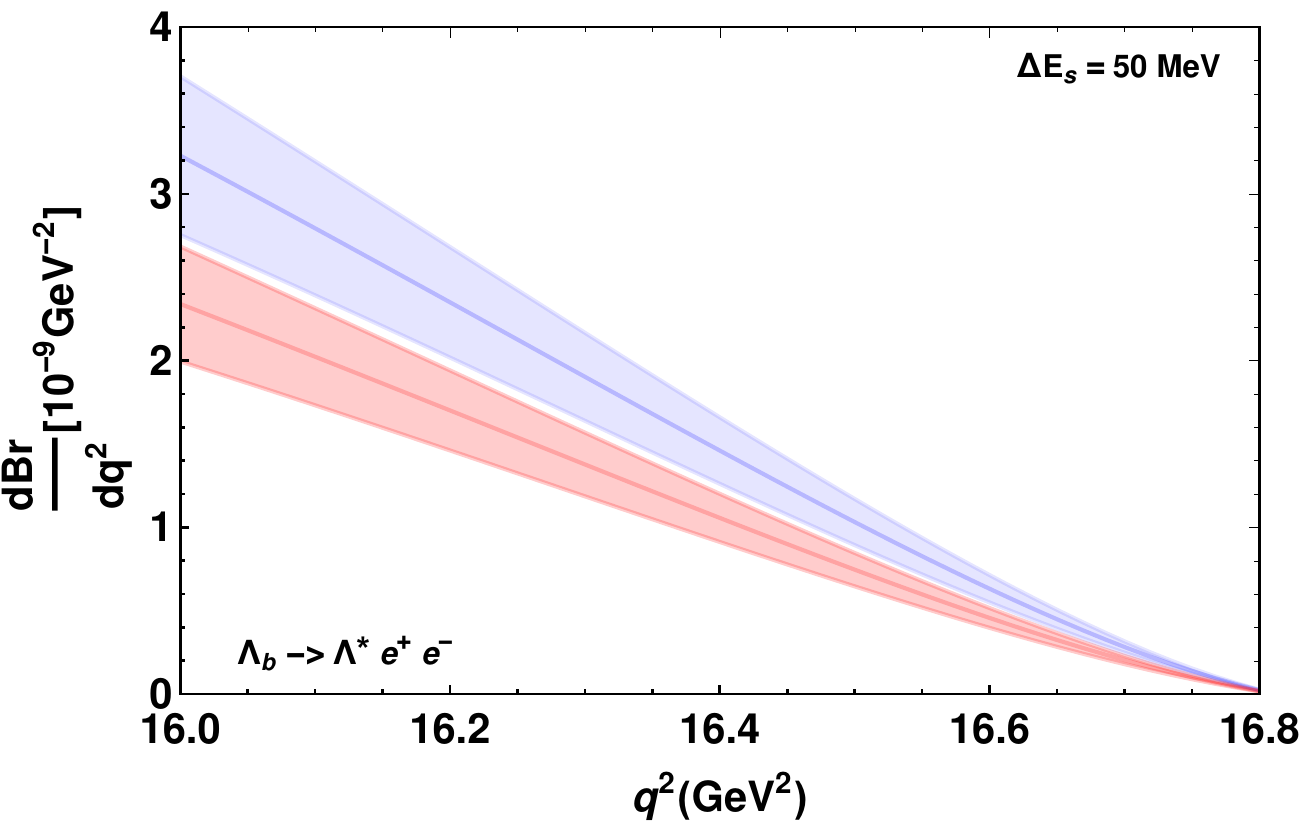}
    \includegraphics[scale=0.33]{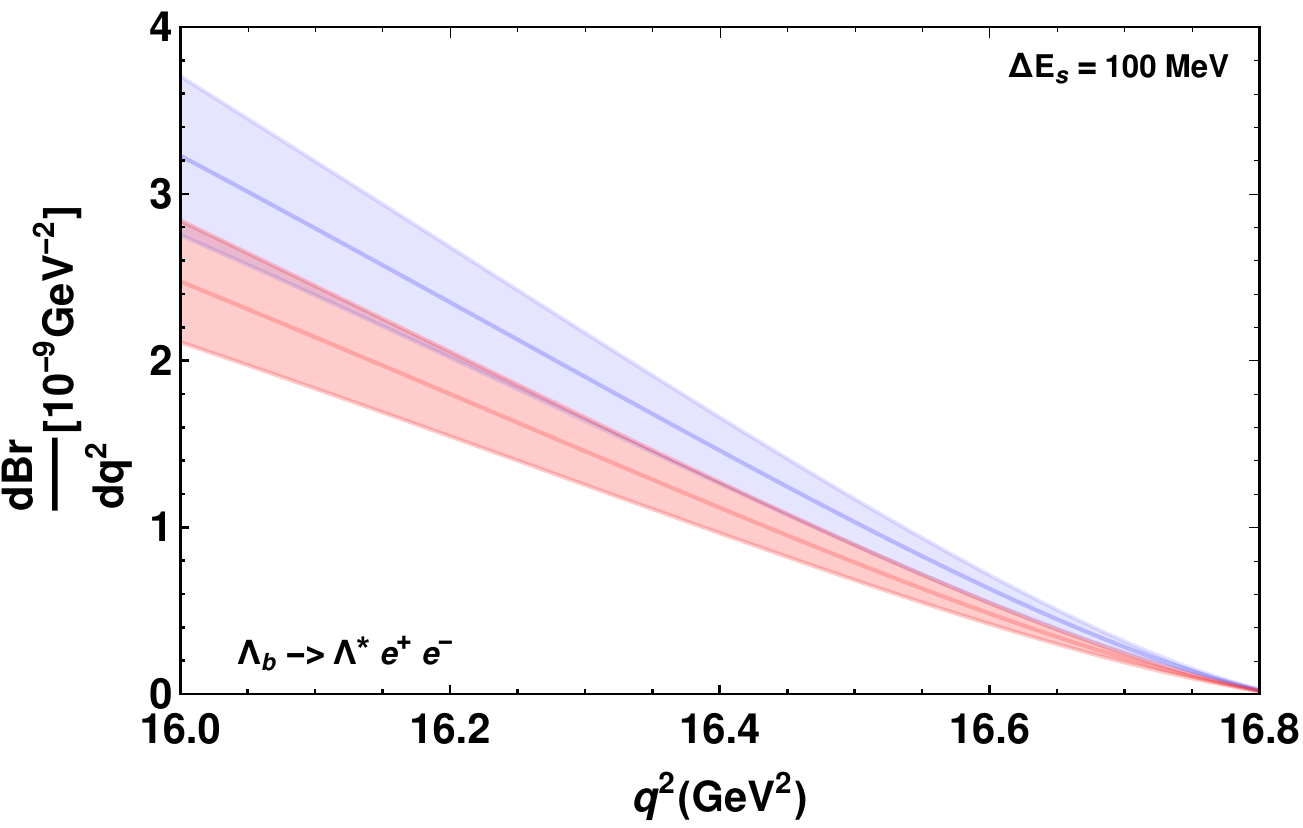}
  \end{center}
  \caption{As in Fig.\,\ref{fig:dBrbkst}, but for $\Lambda_b \to \Lambda^\ast \ell^+ \ell^-$
    instead. The experimental data is taken from ref.\cite{LHCb:2023ptw}.}
  \label{fig:dBrlamdast}
\end{figure}

We now come to the last decay under consideration, namely $\Lambda_b
\to \Lambda^* \ell^+ \ell^-$. With the $\Lambda^*$ being a spin-3/2
particle, the number of possible form factors increases substantially
\cite{Meinel:2021mdj,Meinel:2020owd}, raising the possibility that the theoretical
uncertainties would be even larger in this case. Indeed, to calculate
$\Lambda_b \to \Lambda^\ast$ transition amplitude, in addition to the
statistical uncertainties, once again we use two sets of lattice QCD form factor
parametrization in accordance with ref.~\cite{Meinel:2020owd} so as to
account for systemic uncertainties. The lattice QCD form factors are available only at high-$q^2$ above 16 GeV$^2$. The consequent differential widths are
plotted in Fig.\,\ref{fig:dBrlamdast} as a function of $q^2$. From the
the Fig.\,\ref{fig:dBrlamdast}, we can see that the corrections in the differential branching ratio for
$\Lambda_b \to \Lambda^\ast \ell^+ \ell^-$ is similar to the corrections for
$\Lambda_b \to \Lambda \ell^+ \ell^-$ at high-$q^2$ region.

\section{Summary \label{sec:summary}}
The flavor changing neutral current
transition $b \to s\ell^+\ell^-$, being both loop and CKM
suppressed in the SM, offers a promising ground to search for
tiny effects of new physics (NP). In the past decade,
a series of
experiments testing LFU violation in some such
transitions, had initially indicated a very significant deviation, only to recently settle at $\gsim 1.2 \sigma$ consistency.
However, further investigations of NP are warranted
by the longstanding discrepancies in the differential branching ratios
and the angular observables of $b\to s\ell^+\ell^-$. Unlike LFU
observables, differential branching ratios and angular observables are
plagued by hadronic as well as other theoretical uncertainties. In
this paper, we have assessed the effect of QED corrections to $H_1\to
H_2 \ell^+\ell^-$ where the hadron pair $(H_1, H_2)$ are $(B,
K^{(\ast)})$ and $(\Lambda_b, \Lambda^{(\ast)})$.

With the hadrons being neutral, structure-dependent QED effects generate no new infrared divergences and, for the non-helicity-suppressed transitions considered here, no additional hard-collinear logarithms. They have been neglected, and our results hold in the point-like approximation. As for the rest, at the lowest nontrivial order, the $H_1\to H_2\ell^+\ell^-\gamma$ amplitude splits into a finite part, and a product of non-radiative amplitude and a universal function that is infrared (IR) divergent. Using a fictitious photon mass as an intermediate IR regulator, we explicitly show that all dependences on this mass vanish identically
  when the full set of diagrams are summed over.  The final result
is, however, sensitive to an applied photon momentum cutoff
$\Delta E_s$ below which a photon is too soft
to be detectable in a
detector. While this dependence
  is non-negligible at the NLO level (and expected to be totally
  eliminated only in an all-order calculation), we show that, even at
  the NLL level (achieved by exponentiating the leading order
correction), the dependence is aleady reduced to a great extent.

The overall corrections are negative, leading to a reduction in the partial widths. For $B\to K^\ast\mu^+\mu^-$, this reduction is upto 5\% at low- and upto 11\% in the high-$q^2$ range. For $B\to K^\ast e^+e^-$ the corresponding figures are 15\% and 28\%, respectively. Due to the larger reduction for the electron mode, the LFU ratio $R_{K^\ast}$ is enhanced in comparison to the tree-level result. A similar conclusion is also reached for the $B\to K\ell^+\ell^-$ decays. These enhancements are, however, relative to a non-radiative computation; since the experimental analyses correct for final-state radiation using PHOTOS, they do not imply a tension with the measured $R_K$ and $R_{K^\ast}$, in line with refs.~\cite{Bordone:2016gaq,Isidori:2022bzw}.

For $\Lambda_b\to \Lambda\mu^+\mu^-$ the corrections are up to 5\% at low-$q^2$ and up to 10\% at high-$q^2$; for $\Lambda_b\to \Lambda e^+e^-$, the corresponding numbers are 15\% and 23\% respectively. Once again, the LFU ratio $R_\Lambda$ is larger than the tree level result. Analogous results are obtained for the $\Lambda \to \Lambda^\ast \ell^+\ell^-$ decays as well.

To our knowledge, this is the first determination of these corrections for the baryonic modes $\Lambda_b\to\Lambda^{(\ast)}\ell^+\ell^-$; it provides a point-like baseline for the Monte Carlo modelling of radiation in these channels and for future studies of structure-dependent effects.

\section*{Acknowledgements}
We thank G. Isidori and R. Zwicky for helpful correspondence on
an earlier version of this manuscript.
DC and JD would like to acknowledge Research Grant No.
SERB/CRG/ 004889/SGBKC/2022/04 of the SERB, India. JD also acknowledges the Council of
Scientific and Industrial Research (CSIR), Government of India, for
the SRF fellowship grant with File No. 09/045(1511)/2017-EMR-I. DD would like to thank the DST, Govt. of India for the INSPIRE Faculty Award (grant no. IFA16-PH170) and IIIT Hyderabad for the Seed Grant No. IIIT/R$\&$D Office/Seed-Grant/2021-22/013. 

\appendix
\section{Wilson coeffiecients \label{app:effWCs}}
In addition to the dominant operators $\mathcal{O}_{7,9,10}$, the SM
effective Hamiltonian also receives contributions from current-current
operators $\mathcal{O}_{1-6}$ and the dipole operator
$\mathcal{O}_8$. The values of all the ten Wilson coefficients are listed
  in Table \ref{tab:CCWCs}. As far as the decays in question are concerned, the
contributions of operators $\mathcal{O}_{1-6}$ and $\mathcal{O}_{8}$, can be larged
  subsumed in
$C_{7,9}$ to yield effective coefficients
$C_{7,9}^{\rm eff}$, {\em viz.},
\begin{align}\label{eq: C7eff}
C_7^{\rm eff}   &= C_7 - \frac{1}{3}\left(C_3 + \frac{4}{3}C_4 + 20 C_5 + \frac{80}{3}C_6 \right),\\
%
\label{eq: C9eff}
C_9^{\rm eff} &= C_9 + \frac{4}{3} C_3 + \frac{64}{9}C_5 + \frac{64}{27}C_6
+ h(q^2, 0)\left(-\frac{1}{2}C_3 -\frac{2}{3}C_4 - 8C_5 - \frac{32}{3}C_6\right) \\
& + h(q^2, m_b)\left(-\frac{7}{2}C_3 - \frac{2}{3}C_4 - 38 C_5 - \frac{32}{3} C_6\right) + h(q^2, m_c)\left(\frac{4}{3}C_1 + C_2 + 6 C_3 + 60 C_5\right). \nn\
\end{align}
The functions $h(a,b)$ are given in ref.~\cite{Beneke:2001at}, and the
quark (pole) masses appearing in $C_{7,9}^{\rm eff}$ are $m_b^{\rm pole} = 4.74174$ GeV and
$m_c^{\rm pole} = 1.5953$ GeV \cite{Detmold:2016pkz}. In Table \ref{tab:CCWCs}, we list the values of Wilson coefficients evaluated at $\mu = 4.2 $ GeV.

\begin{table}[ht!]
\begin{center}
\setlength{\tabcolsep}{3pt}
\begin{tabular}{|c | c | c | c | c | c| c| c| c| c|}
\hline \hline
$C_1$ & $C_2$ & $C_3$ & $C_4$ & $C_5$ & $C_6$ & $C_7$ & $C_8$ & $C_9$ & $C_{10}$\\
\hline 
$-0.2877 $ & $1.0101$ & $-0.0060$ & $-0.0860$ & $0.0004$ & $ 0.0011$ & $-0.3361$ & $-0.1821$ & $4.2745$ & $-4.1602$ \\
\hline
\end{tabular}
\end{center}
\caption{The values of Wilson coefficients for $b \to s \ell^+ \ell^- $ FCNC transition at $\mu = 4.2 $ GeV are taken from \cite{Detmold:2016pkz}. \label{tab:CCWCs}}
\end{table}

\section{Analytical expression of $B_{\rm real}$ \label{app:Breal}}
The term $B_{\rm real}$ appearing in
Low's universal soft term (see
eqn.\eqref{eq:intreal} ) is given by
\begin{eqnarray}
B_{\rm real} 
&=& - \frac{1}{4 \pi^2}\int_0^{\Delta E_s} \frac{d^3\vec{k}}{\sqrt{\vec{k}^2 + \lambda^2}} \bigg(\frac{m^2}{(q_1.k)^2} +\frac{m^2}{(q_2.k)^2} -\frac{2 q_1.q_2}{(q_1.k)(q_1.k)} \bigg)\,.
\end{eqnarray}
Since the leptons are on-shell, we may use
\[
  q_1^2 = q_2^2 = m^2 \ , \qquad q^2 = (q_1+q_2)^2.
\]
Being Lorentz-invariant, the integrals are most easily
  calculated in the dilepton rest frame where we have
\[
\vec{q}_1 = - \vec{q}_2 \ , \qquad 
q_1^0 = q_2^0 = \frac{\sqrt{q^2}}{2} \ , \quad {\rm with} \qquad
\beta \equiv \sqrt{1- \frac{4 m^2}{q^2}}
\]
denoting the magnitude of the lepton velocities in this
  frame. Symmetry dictates that the first two of the integrals are
  same, {\em viz.}
\begin{equation}
  \begin{array}{rcl}
    \dis
    \int_0^{\Delta E_s} \frac{d^3\vec{k}}{\sqrt{\vec{k}^2 + \lambda^2}} \frac{1}{(q_1.k)^2} &=& \dis
    \int_0^{\Delta E_s} \frac{d^3\vec{k}}{\sqrt{\vec{k}^2 + \lambda^2}} \frac{1}{(q_2.k)^2} \\[2.5ex]
&=& \dis
    \frac{\pi}{m^2}\left\{ -\ln \left( \frac{\lambda^2}{m^2}\right) + \ln \left(\frac{4\Delta E_s^2}{m^2}\right) \right\} + \frac{1}{\beta}\ln \left( \frac{1-\beta}{1+\beta}\right)\bigg\}.\\
  \end{array}
  \end{equation}
The third integral, on the other hand, is
\begin{eqnarray}
  \int_0^{\Delta E_s} \frac{d^3\vec{k}}{\sqrt{\vec{k}^2 + \lambda^2}}
  \frac{-2 q_1 .q_2}{(q_1.k)(q_2.k)}
  &= & 4\pi\Bigg[
\left(\frac{1+\beta^2}{2\beta} \right)\ln \left(\frac{1-\beta}{1+\beta} \right)\left\{\ln \left(\frac{ 4 \Delta E_s^2}{m^2} \right)- \ln \left( \frac{\lambda^2}{m^2}\right) \right\} \nn\ \\ 
&& \hspace*{1em} +  \left(\frac{1+\beta^2}{2\beta}\right) \left\{ 2 \rm Li_2 \left( \frac{2\beta}{1+\beta}\right)+ \frac{1}{2}\ln^2\left( \frac{1-\beta}{1+\beta}\right) \right\} \Bigg].
\end{eqnarray}

\section{Input parameters}\label{app:input}
The parameters used in this paper, as taken from Particle Data Group-2022 \cite{Workman:2022ynf} and ref.~\cite{Detmold:2016pkz}, are listed below.

\begin{table}[H]
\begin{center}
\setlength{\tabcolsep}{3pt}
\begin{tabular}{|c | c | c | c | c | c| c| c| c| c|}
\hline \hline
$\alpha_e$ & $\alpha_s$ & $|V_{tb}V_{ts}^\ast|$ & $m_b$(GeV) & $m_{\Lambda_b}$(GeV) & $m_{\Lambda}$(GeV) & $m_{\Lambda^\ast}$(GeV)   \\
\hline 
$\frac{1}{133.28} $ & $0.2233$ & $0.0409$ & $4.2$  & $5.6196$ & $ 1.1157$  & 1.520    \\
\hline \hline
  $\tau_{\Lambda_b}$(ps)& $m_B$(GeV) & $m_K$(GeV) & $m_{K^\ast}$(GeV)& $G_F$(GeV$^{-2}$) & $\tau_{B}$(ps)  &  \\
\hline
  $1.466$  &5.2796 & 0.4937 & 0.8958  & $1.1664\times 10^{-5}$& 1.0546&  \\
\hline 
\end{tabular}
\end{center}
\caption{Values of all the input parameters for the relevant processes at $\mu = 4.2 $ GeV. }
\end{table}

\section{Passarino-Veltman functions}\label{app:veltman}
The two point scalar and vector Veltman-Passrino functions are
defined as
\begin{eqnarray}
B_{\left\{0,\mu\right\}}(q_1^2,\lambda^2,m^2) &=& \frac{1}{i \pi^2} \int d^4 k \frac{\{1,k_{\mu}\}}{(k^2 - \lambda^2)\left((k + q_1)^2 - m^2\right)} \ ,
\end{eqnarray}
where we have introduced a non-zero photon mass $\lambda$ as a
  convenient intermediate regulator for the infrared divergences. As
  explained earlier, the final results are entirely independent of
  $\lambda$. Using a ultraviolet cutoff $k_E^2 \leq \Lambda_{\rm
  uv}^2$ (where $k_{E\mu}$ is the Euclideanized momentum),
 the scalar
    integral can be written as
\begin{eqnarray}\label{Veltman}
B_0(q_1^2,\lambda^2,m^2) &=& \int_0^1 dx \Big[- 1 - \ln \Big(\frac{\Delta(x)}{\Lambda_{\rm uv}^2 + \Delta(x)}\Big) + \frac{\Delta(x)}{\Lambda_{\rm uv}^2 + \Delta(x)}\Big] \ ,\label{eq:neglect}
\end{eqnarray}
with
\begin{eqnarray}
\Delta(x) &=& -x(1-x)q_1^2 + m^2 x +(1-x)\lambda^2 \ .
\end{eqnarray} 
With $\Lambda_{\rm uv}$ being much larger than any scale in the
  theory, the last term in eqn.(\ref{Veltman}) may be neglected, and
  we have
\begin{eqnarray}
B_0(m^2,0,m^2) &=& 1 - \ln\Big(\frac{m^2}{\Lambda_{\rm uv}^2}\Big) + \frac{m^2}{3 \Lambda_{\rm uv}^2},
\end{eqnarray}
where we have specialised to the case of on-shell external fermions ($q_1^2 = m^2$). Similarly,
\begin{eqnarray}
B_0^\prime(q_1^2,\lambda^2,m^2) &\equiv&
\frac{d}{dq_1^2}B_0(q_1^2,\lambda^2,m^2)= \int_0^1 dx
\frac{x(1-x)}{\Delta(x)} \ ,
\end{eqnarray}
and, in the on-shell case
\begin{eqnarray}
B_0^\prime(m^2,\lambda^2,m^2) &=& \int_0^1 dx \frac{x(1-x)}{x^2m^2  +(1-x)\lambda^2}	
= - \frac{1}{m^2}- \frac{1}{2m^2}\ln\Big(\frac{\lambda^2}{m^2}\Big).
\end{eqnarray} 
  \\
The Passarino-Veltman functions for the three-point scalar, vector and tensor are defined as follows.
\begin{eqnarray}
C_{\{0,\mu,\mu\nu\}}(q_1,-q_2,\lambda,m, m) &=& \frac{1}{i \pi^2} \int d^4 k \frac{\{1,k_\mu, k_\mu k_\nu\}}{(k^2 - \lambda^2)\left((k + q_1)^2 - m^2\right) \left((k - q_2)^2 - m^2\right)}. \nn\
\end{eqnarray}
In particular, the three-point scalar function is given by
\begin{eqnarray}
C_0(q_1,-q_2,\lambda,m, m) &=& \frac{1}{q^2\beta}\bigg[\ln\left(\frac{\lambda^2}{m^2}\right)\ln\left(\frac{\beta-1}{\beta+1}\right) + 2 \rm Li_2 \left(\frac{\beta-1}{2\beta}\right) +\ln^2      \left(\frac{\beta-1}{2\beta}\right) \nn\ \\ 
&-& \frac{1}{2}\ln^2\left(\frac{\beta-1}{\beta+1}\right) - \frac{\pi^2}{6} \bigg].
%
%
\end{eqnarray}
%

\label{Bibliography}
\bibliographystyle{JHEP}
\bibliography{Refs}
\end{document}